\documentclass[a4paper,11pt,ngerman,UKenglish]{scrartcl}

\usepackage[right=2.6cm, left=2.6cm]{geometry}
\usepackage[utf8x]{inputenc}
\usepackage{amsfonts}
\usepackage{amssymb}
\usepackage{amsmath}
\usepackage{amsthm}
\usepackage{array}
\usepackage{bbm}
\usepackage{booktabs}
\usepackage{caption}
\usepackage{cite}
\usepackage{color}
\usepackage{colortbl}
\usepackage{csquotes}
\usepackage{enumerate}
\usepackage{environ}
\usepackage{fancyhdr}
\usepackage{float}
\usepackage[acronym,toc]{glossaries}
\usepackage{graphicx}
\usepackage{hepunits}
\usepackage{ifpdf}
\usepackage{latexsym}
\usepackage{lscape}
\usepackage{makeidx}
\usepackage{mathtools}
\usepackage{multirow}
\usepackage{placeins}
\usepackage{rotating}
\usepackage{tikz}
\usepackage{units}

\addtokomafont{disposition}{\boldmath} 
\KOMAoptions{bibliography=totoc}
\KOMAoptions{numbers=noendperiod}

\numberwithin{equation}{section}

\newcommand{\M}{\mathcal{M}}
\newcommand{\B}{\mathcal{B}}
\newcommand{\F}{\mathcal{F}}
\newcommand{\order}{\mathcal{O}}
\newcommand{\A}{\mathcal{A}}
\newcommand{\V}{\mathcal{V}}
\newcommand{\R}{\mathcal{R}}
\newcommand{\N}{\mathcal{N}}
\newcommand{\mathH}{\mathcal{H}}
\newcommand{\diff }{{\text{d}}}
\newcommand{\Omnes}{Omn\`{e}s }
\newcommand{\MO}{Muskhelishvili--Omn\`{e}s }
\renewcommand{\vec}[1]{\mathbf{#1}}
\renewcommand{\Re}{\text{Re}\,}
\renewcommand{\Im}{\text{Im}\,}
\newcommand{\cospi}{\cos \theta_{\pi}}

\newcommand{\Bd}{\bar B_{d/s}^0}

\deffootnote{1em}{1em}{\textsuperscript{\thefootnotemark}\ }
\def\bea{\begin{eqnarray}}
\def\eea{\end{eqnarray}}
\def\be{\begin{equation}}
\def\ee{\end{equation}}
\def\bal{\begin{align}}
\def\eal{\end{align}}

\usepackage{booktabs,array}

\newcount\rowc

\usepackage{jheppub_mod}

\title{A model-independent analysis of final-state interactions in \boldmath{$\bar B_{d/s}^0 \to J/\psi \pi\pi $}}

\author[a]{J.~T.~Daub,}
\author[b]{C.~Hanhart,}
\author[a]{and B.~Kubis}

\affiliation[a]{
Helmholtz-Institut f\"ur Strahlen- und Kernphysik (Theorie) and \\ 
Bethe Center for Theoretical Physics, Universit\"at Bonn, 
D-53115 Bonn, Germany}

\affiliation[c]{
Institut f\"ur Kernphysik, Institute for Advanced Simulation, 
and J\"ulich Center for Hadron Physics,
Forschungszentrum J\"ulich, D-52425 J\"{u}lich, Germany}

\emailAdd{daub@hiskp.uni-bonn.de}
\emailAdd{c.hanhart@fz-juelich.de}
\emailAdd{kubis@hiskp.uni-bonn.de}

\abstract{
Exploiting $B$-meson decays for Standard Model tests and beyond requires a precise understanding of the strong final-state interactions
that can be provided model-independently by means of dispersion theory. This formalism allows one to deduce the universal
pion--pion final-state interactions from the accurately known $\pi\pi$ phase shifts and, in the scalar sector, 
a coupled-channel treatment with the kaon--antikaon system.
In this work an analysis of the decays 
$\bar B_d^0 \to J/\psi \pi^+\pi^- $ and $\bar B_s^0 \to J/\psi \pi^+\pi^-$ is presented.
We find very good agreement with the data up to 1.05 GeV in the $\pi\pi$ invariant mass, with a number of parameters reduced significantly
compared to a phenomenological analysis.  
In addition, the phases of the amplitudes are correct by construction, a crucial feature for many $CP$ violation measurements in
heavy-meson decays. 
}

\begin{document}

\maketitle

\section{Introduction}

\begin{figure}
\centering  
\includegraphics[scale=0.8]{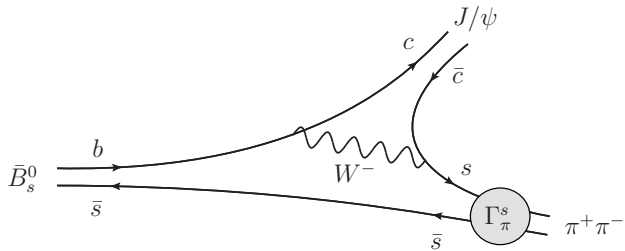}
\caption{The $ \bar B_s^0 \to J/\psi \pi^+\pi^- $ diagram to leading order via $W^-$ exchange. The hadronization into pions ($S$-wave dominated) proceeds through the pion strange scalar form factor $\Gamma^s_\pi(s)$. In the case of the $ \bar B_d^0  \to J/\psi \pi^+\pi^- $ decay, with $s \leftrightarrow d$, the pions are generated out of a non-strange scalar source, i.e.\ $\Gamma^s_\pi(s)$ is replaced by the pion non-strange scalar form factor $\Gamma^n_\pi(s)$ for $S$-wave and by the vector form factor for $P$-wave pions.}
\label{fig:graph}
\end{figure}

$B$-meson decays can be exploited for Standard Model tests and beyond, in particular to determine the Cabibbo--Kobayashi--Maskawa (CKM) couplings and to study $CP$ violation. 
For a theoretical description of many of these decays, 
it is mandatory to understand the strong final-state interactions in terms of amplitude
analysis techniques~\cite{Battaglieri:2014gca}, with tight control over the magnitudes and phase motions of the 
various partial waves involved.
For example, the decays $B \to f_0(980) K_S$ and $B \to \phi(1020) K_S$ are explored for an experimental determination of the $CP$ asymmetry $\sin 2\beta$~\cite{Dalseno:2008wwa,Aubert:2009me,Nakahama:2010nj,Lees:2012kxa}, $\beta$ being one of the angles of the unitarity triangle, which requires precise knowledge of the strange and non-strange scalar form factors that we discuss in this article.
We focus on the decays $\bar B_d^0 \to J/\psi \pi^+\pi^- $ and $\bar B_s^0 \to J/\psi \pi^+\pi^- $, measured by the LHCb collaboration~\cite{Aaij:2014siy,Aaij:2014emv}.
The tree-level process of the weak decay into $J/\psi$ and a $q \bar q$ pair is depicted in Fig.~\ref{fig:graph} (exemplarily for the $\bar B_s^0$ decay). 
These analyses complement former related studies of $\bar B_d^0$ and $\bar B_s^0$ decays by the BaBar~\cite{Aubert:2002vb}, Belle~\cite{Li:2011pg}, CDF~\cite{Aaltonen:2011nk}, 
and D0~\cite{Abazov:2011hv} Collaborations as well as older LHCb results~\cite{Aaij:2013zpt, LHCb:2012ae}.
Universality of final-state interactions dictates that
the hadronization into pions and the rescattering effects in the $\pi^+ \pi^-$ system for $S$- and $P$-waves are closely related to the scalar and vector pion form factors, respectively. We describe these form factors using dispersion theory, using \Omnes (or Muskhelishvili--Omn\`{e}s) representations. 
In doing so we exploit the fact that LHCb found no obvious structures in the $J/\psi \pi^+$ invariant mass distribution, suggesting that left-hand-cut contributions in the $\pi^+\pi^-$ system due to the crossed-channel $J/\psi \pi^+$ interaction are small and can be neglected.

The advantage of the dispersive framework is that all constraints imposed by
analyticity (i.e., causality) and unitarity (probability conservation) are fulfilled by construction.
Further, it is a model-independent approach, so we do not have to specify any contributing resonances or conceivable non-resonant backgrounds.
For the vector form factor a single-channel (elastic) treatment works very well below 1~\GeV. In the scalar sector the strong coupling of two $S$-wave pions to $K\bar K$ near 1~\GeV\ due to the $f_0(980)$ resonance, causing a sharp onset of the $K\bar K$ inelasticity, necessitates a coupled-channel treatment. Therefore a two-channel \MO problem is solved.
This two-channel approach breaks down at energies where inelasticities caused by $4\pi$ states become important, we are thus not able to cover the complete phase space, but restrict ourselves to the low-energy range $\sqrt{s} \le 1.05~\GeV$.

In Ref.~\cite{Aaij:2014siy} the $\bar B_d^0$ decay is described by six resonances in the $\pi^+ \pi^-$ channel, $f_0(500)$, $\rho(770)$, $\omega(782)$, $ \rho(1450)$, $\rho(1700)$, and $f_2(1270)$,  which are modeled by Breit--Wigner functions. 
This parametrization of especially the $f_0(500)$ meson is somewhat precarious, as the broad bump structure of this scalar resonance is not well described by a Breit--Wigner shape.  As demonstrated for the first time in the context of $B$ decays
in Ref.~\cite{Gardner:2001gc}, it should be replaced by the corresponding scalar form factor.
In the present work this idea is extended and rigorously applied using form factors derived from dispersion theory.
In particular, there is no need to parametrize any resonance, since the input required to describe the final-state interactions 
is taken from known phase shifts, and therefore the $f_0(500)$ appears naturally in the non-strange scalar form factor.
The $\bar B_s^0$ decay, described in the experimental analysis by five resonances, $f_0(980)$, $f_0(1500)$, $ f_0(1790)$, $f_2(1270)$, and $f_2'(1525)$ (Solution~I) or with an additional non-resonant contribution (Solution~II), dominantly occurs in an $S$-wave state~\cite{Aaij:2014emv}, while the $P$-wave is shown to be negligible. Given the almost pure $\bar ss$ source the pions are generated from, this decay shows great promise to provide insight into the strange scalar  form factor. 

The idea of such a \enquote{scalar-source model}, where an $S$-wave pion pair is generated out of a quark--antiquark pair and the final-state interactions are described by the scalar form factor, is also used in Ref.~\cite{Liang:2014tia} for the description of the $\bar B_s^0$ and $\bar B_d^0$ decays into the scalar resonances $f_0(980)$ and $f_0(500)$, respectively.
It was employed earlier e.g.\ in analyses of the decay of the $J/\psi$ into a vector meson ($\omega$ or $\phi$) and a pair of pseudoscalars ($\pi\pi$ or $K\bar K$)~\cite{Meissner:2000bc, Lahde:2006wr}. In these references the strong-interaction part is described by a chiral unitary theory including coupled channels, which yields a dynamical generation of the scalar mesons.  In contrast to the present study, the very precise information available on 
pion--pion~\cite{Ananthanarayan:2000ht,GarciaMartin:2011cn,CCL2012,CCLprep} and pion--kaon~\cite{BDM04} 
phase shifts is not strictly implemented there.
Related studies using the chiral unitary approach are performed in Ref.~\cite{Bayar:2014qha}, where the $J/\psi$--vector-meson final state is analyzed, and in Ref.~\cite{Xie:2014gla}, which includes resonances beyond 1~\GeV.
In contrast to models of dynamical resonance generation, the scalar resonances are considered as $q \bar q$ or tetraquark states in Ref.~\cite{Stone:2013eaa}.
Other theoretical approaches employ light-cone QCD sum rules to describe the form factors~\cite{Colangelo:2010bg}.
Progress on the short-distance level is made in Ref.~\cite{Wang:2015uea}, where the factorization formulae (which we treat in a naive way) are improved in a perturbative QCD framework.

This manuscript is organized as follows. 
In Sec.~\ref{sec:amplconstr}, we review the construction of the transversity amplitudes and partial waves, after sketching the kinematics. We provide 
explicit expressions that relate the theoretical quantities to the angular moments determined in experiment.
Section~\ref{sec:omnes} is focused on the \Omnes formalism. 
The fits to the LHCb data, using the $\Bd \to J/\psi \pi^+ \pi^-$ angular moment distributions, are discussed in Sec.~\ref{sec:fits}, where we use several configurations with and without $D$-wave corrections to study the impact of certain corrections to our fits. 
We also predict the $S$-wave amplitude for the related $\bar B_s^0 \to J/\psi K^+  K^- $ decay.
The paper ends with a summary and an outlook in Sec.~\ref{sec:summary}.
Some technical details are relegated to the appendices.


\section{Kinematics, decay rate, and angular moments}\label{sec:amplconstr}

In this section we derive the decay rate and angular moments for the $\bar B_d^0 \to J/\psi \pi^+\pi^-$ decay mode in terms of partial-wave amplitudes up to $D$-waves, employing the transversity formalism of Ref.~\cite{Faller:2013dwa}. The formalism works analogously for the $\bar B_s^0$ decay.

\subsection{Kinematics}\label{sec:kinematics}

\begin{figure}
\centering  
\includegraphics[scale=0.6]{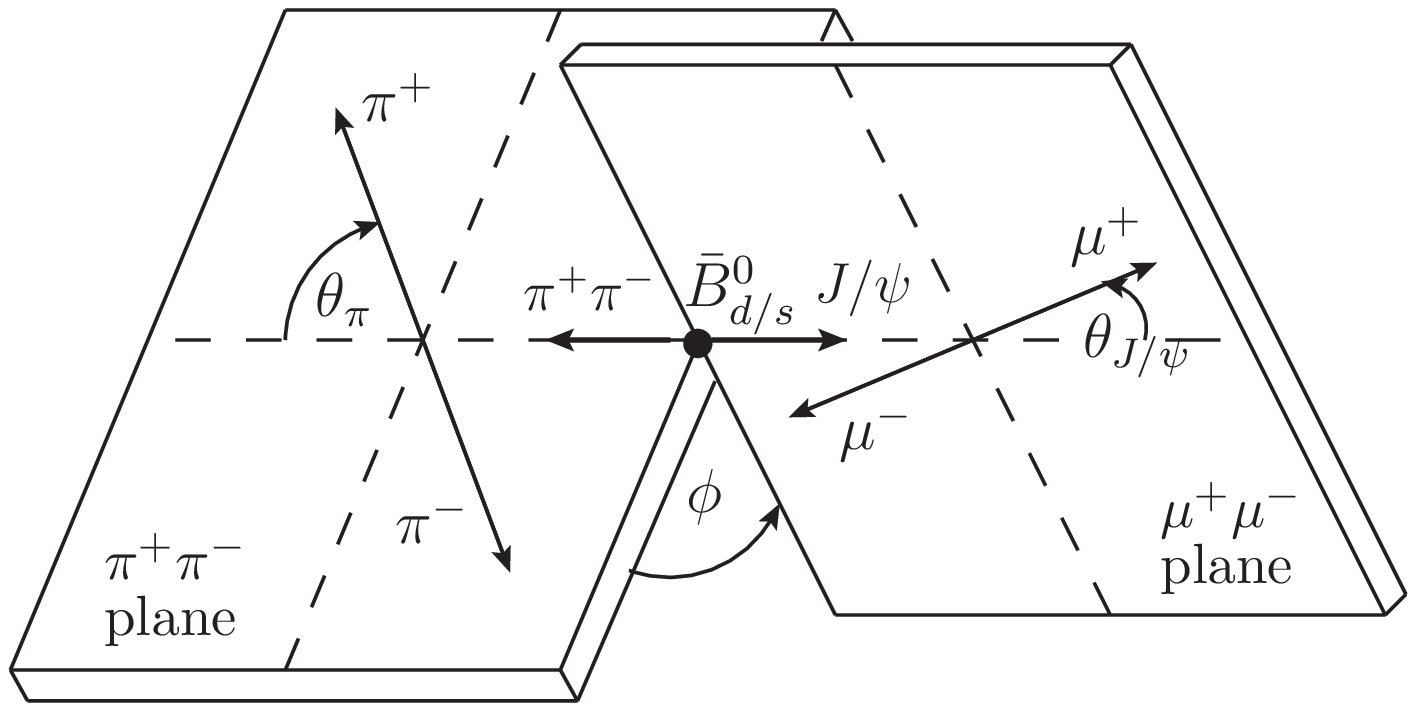}
\caption{Definition of the kinematical variables for $\Bd \to J/\psi \pi^+\pi^- $.}
\label{fig:kinematics}
\end{figure}

The kinematics of the decay $ \Bd (p_B) \to J/\psi (p_\psi) \pi^+ (p_1) \pi^- (p_2) $ ($J/\psi \to \mu^+ \mu^-$) can be described by four variables: 
\begin{itemize} 
\item the invariant dimeson mass squared, $s=(p_1 + p_2)^2$, 
\end{itemize}
and three helicity angles, see Fig.~\ref{fig:kinematics},
\begin{itemize}  
\item $\theta_{J/\psi}$, the angle between the $\mu^+$ in the $J/\psi$ rest frame ($\Sigma_{J/\psi}$) and the $J/\psi$ in the $\Bd$ rest frame ($\Sigma_B$);
\item $\theta_\pi$, the angle between the $\pi^+$ in the $\pi^+\pi^-$ center-of-mass frame $\Sigma_{\pi\pi}$ and the dipion line-of-flight in $\Sigma_B$;
\item $\phi$, the angle between the dipion and the dimuon planes, where the latter originate from the decay of the $J/\psi$.
\end{itemize}
The three-momenta of either of the two pions in the dipion center-of-mass system ($\vec p_\pi$) and of the $J/\psi$ in the $\Bd$ rest frame ($\vec p_\psi$) are given by 
\be
|\vec p_\pi |=\frac{\lambda^{1/2}(s,M_\pi^2,M_\pi^2)}{2\sqrt{s}} \equiv \frac{\sigma_\pi \sqrt{s}}{2}, \quad
|\vec p_\psi |=\frac{\lambda^{1/2}(s,m_\psi^2,m_B^2)}{2 m_B} \equiv \frac{X}{m_B},
\ee
with the  Käll\'{e}n function $\lambda(a,b,c) = a^2+b^2+c^2 - 2ab -2ac - 2bc$.

We define the two remaining Mandelstam variables as
\be
t = (p_B - p_1)^2 \quad {\rm and} \quad u = (p_B - p_2)^2,
\ee
where the difference of these two determines the scattering angle $\theta_\pi$,
\be
t - u = - 2 p_\psi (p_1 -p_2) = - 2 \sigma_\pi X \cos \theta_\pi .
\ee
Further, we introduce two additional vectors as combinations of the above four-momenta,
\be\label{eq:PQ}
P^\mu = p_1^\mu + p_2^\mu, \quad Q^\mu = p_1^\mu - p_2^\mu.
\ee

\subsection{Matrix element}\label{sec:matrixelement}

To calculate the matrix element we make use of the effective Hamiltonian that governs the $b \to c \bar c d$ transition~\cite{Buchalla:1995vs},
\be\label{eq:Heff}
\mathcal{H}_{\rm eff} = \frac{G_F}{\sqrt{2}} \big\{ V_{cb} V^*_{cd} \left[ C_1(\mu) O_1(\mu) + C_2(\mu) O_2(\mu)  \right] + \dots \big\},
\ee
where the $C_i$ are Wilson coefficients and the $O_i$ local current--current operators
\bea
O_1 &=& \bar c_k \gamma_\mu (1 - \gamma_5) b_l \,\, \bar d_l \gamma^\mu (1 - \gamma_5) c_k =  \bar c_k \gamma_\mu (1 - \gamma_5) c_k \,\, \bar d_l \gamma^\mu (1 - \gamma_5) b_l, \nonumber\\
O_2 &=& \bar c_k \gamma_\mu (1 - \gamma_5) b_k \,\, \bar d_l \gamma^\mu (1 - \gamma_5) c_l = \bar c_k \gamma_\mu (1 - \gamma_5) c_l \,\, \bar d_l \gamma^\mu (1 - \gamma_5) b_k,
\eea
with $k,l$ being color indices.
In the second step the quark operators are regrouped by means of a Fierz rearrangement.
The ellipses in $\mathcal{H}_{\rm eff}$ denote operators beyond tree-level, including penguin topologies.
$V_{cb}$ and $V_{cd}$ are the CKM matrix elements for $c \to b$ and $c \to d $ (where $ V_{cd}$ is to be replaced by  $V_{cs}$ for the $\bar B_s^0$ decay),  and $G_F = 1.166365 \times 10^{-5}~\GeV^{-2}$ is the Fermi constant.

Under the assumption that the final-state interaction between the $J/\psi$ and the pions is negligible (no obvious structures are found in the $J/\psi \pi$ channel experimentally~\cite{Aaij:2014siy, Aaij:2014emv}, and a close-to-zero $J/\psi \pi$ scattering length $a_{J/\psi \pi} = -0.01(1)$~fm results from lattice calculations~\cite{Liu:2008rza}) a factorization approach appears to be justified. Note that on the quark level this naive factorization ansatz may be spoiled~\cite{Beneke:2000ry,Diehl:2001xe}, 
for instance due to (large) penguin contributions that we have neglected in Eq.~\eqref{eq:Heff}~\cite{Aaij:2014vda,Frings:2015eva}. 
However, a more complicated structure of the source term does not conflict with our approach: 
any factorization limitations due to color structures do not concern the \textit{hadronic} final-state interaction,
for which the short-distance factorizations are sufficient but not mandatory. All we use is the fact that the $B$ decays provide clean $\bar q q$ sources of much shorter range than that of the final-state interaction. In our approach, any deviations from clean point sources would be parametrized by derivatives of the source term. An excellent fit to the data even without those correction terms is a proof that with respect to the final-state interactions the sources can be regarded as point-like.

We express the matrix elements of the four-quark operators by two independent hadronic currents, valid if the $c \bar c$ system produced by the hadronization of the virtual $W^-$ is well separated from the spectator quark system. For the decay of the $\bar B_d^0$ meson considered here the matrix element is, 
in analogy to the $\bar B_s^0$ expression given in Ref.~\cite{Colangelo:2010wg}, written as
\bea\label{eq:Mfi}
\M_{fi} &=& \frac{G_F}{\sqrt{2}} V_{cb} V_{cd}^*\, a^{\rm eff} (\mu)\underbrace{\langle \pi^+(p_1) \pi^-(p_2) | \bar{d} \gamma_\mu (1-\gamma_5) b | \Bd (p_B) \rangle}_{\M^{\pi\pi}_\mu} \times \underbrace{\langle J/\psi(p_\psi, \epsilon) | \bar{c} \gamma^\mu c | 0 \rangle}_{\M^{c \bar c\, \mu}},\nonumber\\
\M^{\pi\pi}_\mu &=&  \frac{M_\psi P_{(0)}^{\mu}}{X}  \F_0 + \frac{Q_{(\parallel)}^\mu}{\sqrt{s}}\F_\parallel - \frac{i \bar p_\perp^\mu}{\sqrt{s}} \F_\perp, \qquad
\M^{c \bar c}_\mu = f_\psi M_\psi \epsilon_\mu^* (p_\psi,\lambda),
\eea
with $a^{\rm eff} = C_1(\mu) + C_2(\mu)/N_c + \dots$, the ellipses denoting combinations of Wilson coefficients due to penguin diagrams, we have not taken into account explicitly.

The scale ($\mu$) dependence of the Wilson coefficients is cancelled by the scale dependence of the hadronic matrix elements, cf.\ Sec.~\ref{sec:omnes}; $\mu$ is chosen to be of order $\order(m_B)$, such that heavier particles, in particular the $W$, are integrated out.

The current that creates the $J/\psi$ from the vacuum is related to the decay constant $f_\psi$. The matrix element containing the pions is given by the three transversity form factors $\F_0$, $\F_\parallel$, and $\F_\perp$, corresponding to the orthogonal basis of momentum vectors~\cite{Faller:2013dwa} 
\be\label{eq:newbasis}
p_\psi^\mu, \quad
\bar p_{(\perp)}^\mu = \frac{\epsilon^{\mu\alpha\beta\gamma}}{X} (p_\psi)_\alpha P_\beta Q_\gamma, \quad
Q_{(\parallel)}^\mu = Q^\mu -\frac{(P \cdot p_\psi)(Q \cdot p_\psi)}{X^2} P^\mu + \frac{s (Q \cdot p_\psi)}{X^2} p_\psi^\mu.
\ee
We define $\epsilon_{\mu\nu\rho\sigma}$ such that $\epsilon_{0123} = - \epsilon^{0123} = +1$.

The partial-wave expansions of the transversity form factors read\footnote{Though we expect the $D$- and higher waves to be small and therefore describe only $S$- and $P$-waves in the \Omnes formalism, we present the formulae including the $D$-wave contribution, as we will study their impact at a later stage. }
\allowdisplaybreaks{
\begin{align}
\F_0 (s,\theta_\pi) &= \sum_{\ell} \sqrt{2 \ell + 1}\, \F_0^{(\ell)} (s) P_\ell(\cos\theta_\pi) \nonumber\\
&= \F_0^{(S)} (s) + \sqrt{3} \cos\theta_\pi \F_0^{(P)} (s)  + \frac{\sqrt{5}}{2} \left(3 \cos^2\theta_\pi - 1 \right) \F_0^{(D)} (s) + \dots \,,\nonumber\\
\F_{\parallel,\perp} (s,\theta_\pi) &= \sum_{\ell} \frac{\sqrt{2 \ell + 1}}{\sqrt{\ell (\ell + 1)}} \F_{\parallel,\perp}^{(\ell)} (s) P_\ell'(\cos\theta_\pi) = \sqrt{\frac{3}{2}} \left( \F_{\parallel,\perp}^{(P)}(s) + \sqrt{5} \cos \theta_\pi \F_{\parallel,\perp}^{(D)}(s) \right) + \dots, \label{eq:F-PWE}
\end{align}
}%
where the ellipses denote waves larger than $D$-waves.
In Appendix~\ref{app:FFandPWE} the relation to the helicity form factors is briefly sketched, which have a well-known partial-wave expansion.

\subsection{Decay rate and angular moments}

When comparing the angular moments to the experimental data we have to deal with flavor-averaged expressions due to the $B^0$--$\bar B^0$ mixing and take into account the $CP$-conjugated amplitudes (the $B_d^0$ decay mode) as well. 
Since the interfering term between the amplitudes is negligibly small~\cite{Aaij:2014siy}, the decay rate can be written as the sum of the decay rates for the direct $\bar B_d^0$ and the mixed $CP$-conjugated $B_d^0$ mode,
\be\label{eq:decayratesum}
\frac{\diff^2\Gamma\left(\bar B_d^0 \to J/\psi \pi^+\pi^-\right)}{\diff\sqrt{s} \,\diff\cos\theta_\pi} \approx \frac{\diff^2\Gamma\left({\rm direct}\right)}{\diff\sqrt{s} \,\diff\cos\theta_\pi}  + \frac{\diff^2\Gamma\left(B_d^0 \to J/\psi \pi^+\pi^-\right)}{\diff\sqrt{s} \,\diff\cos\theta_\pi} .
\ee
Note that this neglect is less justified when applying the formulae to the $\bar B_s^0$ decay rate. In the analysis of Ref.~\cite{Aaij:2014emv} an interference term is added to Eq.~\eqref{eq:decayratesum}. However, in Sec.~\ref{sec:results2} we find that it is sufficient to take into account $S$-waves. In that case the
interference term does not affect the fit procedure and merely generates a tiny shift of the resulting
fit parameter (the normalization $c_0^s$).

In this section we provide expressions for one particular mode. The $CP$-related amplitude can be deduced straightforwardly by multiplying the transversity partial-wave amplitudes with $CP$ eigenvalues as outlined in detail below (cf.\ the discussion around Eq.~\eqref{eq:cprel}).

The differential decay rate is given by
\begin{align}
\frac{\diff^2\Gamma}{\diff\sqrt{s} \,\diff\cos\theta_\pi} &= \frac{G_F^2 |V_{cb}|^2 |V_{cd}|^2 f_\psi^2 M_\psi^2 X \sigma_\pi \sqrt{s}}{4 (4\pi)^3 m_B^3}  \nonumber\\
&\times\bigg\{ \Big|\F_0^{(S)} (s) + \sqrt{3} \cos\theta_\pi \F_0^{(P)} (s)  + \frac{\sqrt{5}}{2} \left(3 \cos^2\theta_\pi - 1 \right) \F_0^{(D)} (s)  \Big|^2 \nonumber\\
& +\frac{3}{2} \sigma_\pi^2 \sin^2 \theta_\pi\bigg(\Big|\F_\parallel^{(P)} + \sqrt{5} \cos \theta_\pi \F_{\parallel}^{(D)}(s) \Big|^2 +  \Big|\F_\perp^{(P)} + \sqrt{5} \cos \theta_\pi \F_{\perp}^{(D)}(s) \Big|^2 \bigg) \bigg\}, \label{eq:d2Gamma}
\end{align}
see Appendix~\ref{app:FFandPWE} for details.
By weighting this decay rate by spherical harmonic functions $Y_l^0(\cospi)$, we define the angular moments
\be
\langle Y_l ^0 \rangle (s)= \int_{-1}^1 \frac{\diff^2\Gamma}{\diff\sqrt{s}\, \diff\cospi}  Y_l ^0 (\cospi) \diff\cospi.
\ee
With the orthogonality property
\be
\int_{-1}^1 Y_i^0 (\cospi) Y_j^0(\cospi) \diff\cospi = \frac{\delta_{ij}}{2\pi},
\ee
we obtain
{\allowdisplaybreaks
\begin{align}\label{eq:Y}
\sqrt{4 \pi} \langle Y _0 ^0 \rangle &=  \frac{G_F^2 |V_{cb}|^2 |V_{cd}|^2 f_\psi^2 M_\psi^2 X \sigma_\pi \sqrt{s}}{2 (4\pi)^3 m_B^3} \bigg\{ \Big|\F_0^{(S)} \Big|^2 + \Big|\F_0^{(P)}  \Big|^2 +  \Big|\F_0^{(D)} \Big|^2 \nonumber\\
&\quad  +  \sigma_\pi^2 \bigg(\Big|\F_\parallel^{(P)}\Big|^2 +  \Big|\F_\perp^{(P)} \Big|^2 + \Big|\F_\parallel^{(D)}\Big|^2 +  \Big|\F_\perp^{(D)} \Big|^2\bigg)\bigg\}, \nonumber\\
\sqrt{4 \pi} \langle Y _1 ^0 \rangle &=
\frac{G_F^2 |V_{cb}|^2 |V_{cd}|^2 f_\psi^2 M_\psi^2 X \sigma_\pi \sqrt{s}}{ (4\pi)^3 m_B^3} \bigg\{  \Re \Big(\F_0^{(S)}  \F_0^{(P)^*}  \Big) + \frac{2}{\sqrt{5}}  \Re \Big(\F_0^{(P)}  \F_0^{(D)^*}  \Big)  \nonumber\\
&\quad  + \sqrt{\frac{3}{5}} \sigma_\pi^2 \bigg[ \Re \Big(\F_\perp^{(P)}  \F_\perp^{(D)^*}\Big)  +  \Re \Big(\F_\parallel^{(P)}  \F_\parallel^{(D)^*}  \Big)\bigg] \bigg\}, \nonumber\\
\sqrt{4 \pi} \langle Y _2 ^0 \rangle &= 
\frac{G_F^2 |V_{cb}|^2 |V_{cd}|^2 f_\psi^2 M_\psi^2 X \sigma_\pi \sqrt{s}}{(4\pi)^3 m_B^3}  \bigg\{ 
\Re \Big(\F_0^{(S)}  \F_0^{(D)^*} \Big) +
\frac{1}{\sqrt{5}}\Big|\F_0^{(P)} \Big|^2 +  \frac{\sqrt{5}}{7}\Big|\F_0^{(D)} \Big|^2  \nonumber\\
&\quad    - \frac{\sigma_\pi^2}{2\sqrt{5}}   \bigg(\Big|\F_\parallel^{(P)}\Big|^2 +  \Big|\F_\perp^{(P)} \Big|^2 \bigg) 
+ \frac{\sigma_\pi^2 \sqrt{5}}{14}   \bigg(\Big|\F_\parallel^{(D)}\Big|^2 +  \Big|\F_\perp^{(D)} \Big|^2 \bigg)\bigg\},
\end{align}
}%
where $\langle Y _0 ^0 \rangle$ corresponds to the event distribution, $\langle Y _1 ^0 \rangle$ describes the interference between $S$- and $P$-wave as well as $P$- and $D$-wave amplitudes, and $\langle Y _2 ^0 \rangle$ contains $P$-wave, $D$-wave, and $S$--$D$-wave interference contributions. 

The corresponding expressions for the $CP$-conjugated modes are related to the above equations by certain sign changes due to the $CP$ eigenvalues $\eta_{CP} = \pm 1$ in the definitions of the transversity partial-wave amplitudes, as already mentioned in the beginning of this section. We declare the amplitudes $\F_\tau^{(\ell)}$ to describe the $B_d^0$ decay, then the corresponding $\bar B_d^0$ decay amplitudes are given by
\be
\label{eq:cprel}
\bar \F_\tau^{(\ell)} = \eta_{CP} \F_\tau^{(\ell)},
\ee
with $\eta_{CP} = + 1$ for the $\tau = 0, \parallel$ $P$-waves and the $\tau = \perp$ $D$-wave, and $\eta_{CP} = - 1$ otherwise.
Consequently the angular moments $\langle Y _0 ^0 \rangle$ and $\langle Y _2 ^0 \rangle$ are unchanged under $CP$ conjugation, while the conjugated moment $\langle Y _1 ^0 \rangle$ has opposite sign, such that when considering flavor-averaged quantities and summing over the $B_d^0$ and $\bar B_d^0$ contributions, $\langle Y _1 ^0 \rangle$ vanishes.
In the following we thus consider $\langle Y _0 ^0 \rangle$ and $\langle Y _2 ^0 \rangle$ only. 

\section{\Omnes formalism}\label{sec:omnes}

We describe the $S$- and $P$-wave amplitudes using dispersion theory. This approach allows us to treat the pion--pion rescattering effects in a model-independent way, based on the fundamental principles of unitarity and analyticity: the partial waves are analytic functions in the whole $s$-plane except for a branch-cut structure dictated by unitarity. In the following we deal with the functions $f_\ell^I(s)$ (referring to isospin $I$ and angular momentum $\ell$) that possess a right-hand cut starting at the pion--pion threshold $s_{\rm thr} = 4M_\pi^2$ and are analytic elsewhere, i.e.\ we do not consider any left-hand-cut or pole structure related to crossing symmetry. This is justified from the observation that there are practically no structures observed for the crossed $J/\psi \pi^+$ channel in the region of interest~\cite{Aaij:2014siy}.

Considering two-pion intermediate states only, Watson's theorem holds, i.e.\ the phase of the partial wave is given by the elastic pion--pion phase shift~\cite{Watson:1954uc}, and the discontinuity across the cut can be written as
\be\label{eq:sDR}
{\rm disc} f_\ell^I(s) = f_\ell^I(s+i\epsilon) -f_\ell^I(s-i\epsilon) = 2i \sigma_\pi f_\ell^I (s) \left[ t_\ell^I (s) \right]^* = f_\ell^I(s) e^{-i \delta_\ell^I} \sin \delta_\ell^I.
\ee
A solution of this unitarity relation can be constructed analytically, setting (compare Ref.~\cite{Heyn:1980bh})
\be
f_\ell^I(s) = P(s)  \Omega_\ell^I (s), \label{eq:P.O}
\ee
where $P(s)$ is a polynomial not fixed by unitarity, 
and the \Omnes function $\Omega_\ell^I(s)$ is entirely determined by the phase shift $\delta_\ell^I(s)$~\cite{Omnes:1958hv},
\be
\Omega_\ell^I (s) = {\rm exp} \left\{\frac{s}{\pi}  \int_{s_{\rm thr}}^{\infty} \frac{\delta_\ell^I(s')}{s'(s'-s-i \epsilon)}\diff s' \right\},
\ee
with
\be
\Omega_\ell^I(0) = 1 \quad {\rm and } \quad \Omega_\ell^I(s) \neq 0 \quad \forall \; s.
\ee 

The $P$-wave amplitudes can be well described in the elastic approximation up to energies of roughly 1~\GeV.\footnote{In the following we will suppress the isospin indices as Bose symmetry demands the $S$-waves to be isoscalar, while the $P$-waves are restricted to $I=1$.} The simplest possible application is the pion vector form factor $F_\pi^V(s)$,
\be
\langle 0 | j_{\rm em}^\mu(0) | \pi^+(p_1)\pi^-(p_2) \rangle =  (p_2-p_1)^\mu F_\pi^V(s),
\quad j_{\rm em}^\mu = \frac{2}{3} \bar u\gamma^\mu u - \frac{1}{3} \bar d\gamma^\mu d,
\label{eq:current}
\ee
which obeys a representation like \eqref{eq:P.O} with a linear polynomial $P_{F_\pi^V}(s) = 1+\alpha s$, $\alpha \approx 
0.1~\GeV^{-2}$~\cite{Hanhart:2013vba} up to $\sqrt{s} \approx 1~\GeV$, with the exception of a small energy region
around the $\omega$ resonance that couples to the two-pion channel
via isospin-violating interactions.  In this context it is important to note that the electromagnetic current $j_{\rm em}^\mu$,
introduced in Eq.~\eqref{eq:current}, can be decomposed as
\be \label{eq:em-iso}
j_{\rm em}^\mu = \frac{1}{2} \big( \bar u\gamma^\mu u - \bar d\gamma^\mu d\big)
+ \frac{1}{6} \big( \bar u\gamma^\mu u + \bar d\gamma^\mu d\big).
\ee
Thus it contains with the first term an isovector and with the second term an isoscalar component.
The latter couples directly to the $\omega$, whose decay into $\pi^+\pi^-$ is suppressed by isospin,
but enhanced by a small energy denominator (i.e., the small width of the $\omega$), hence leading to a clearly observable
effect in the pion form factor~\cite{Aubert:2009ad,Ambrosino:2010bv,Ablikim:2015lsa}.  Theoretically, this effect
is correctly taken into account by the replacement~\cite{Gardner:1997ie,Leutwyler:2002hm,Hanhart:2012wi}
\be \label{eq:omegamix}
P_{F_\pi^V}(s) \Omega_1^1 (s) \longrightarrow P_{F_\pi^V}(s) \Omega_1^1(s) \bigg( 1+\frac{\kappa_{\rm em} \, s}{M_\omega^2 -iM_\omega\Gamma_\omega - s} \bigg) .
\ee
Note that in case of the $\omega$ the use of a Breit--Wigner parametrization is appropriate since
the $\omega$ pole is located far above the relevant decay thresholds and since $\Gamma_\omega=8.5$~\MeV\ 
is very small.
A fit of the form factor parametrization introduced in Eq.~\eqref{eq:omegamix} 
to the KLOE data~\cite{Ambrosino:2010bv} yields $\kappa_{\rm em} \approx 1.8\times 10^{-3}$.
This fixes the strength of the so-called $\rho$--$\omega$ mixing amplitude phenomenologically.
The isospin-violating coupling $\kappa_{\rm em}$ is of the usual size, however, near the $\omega$ peak
its smallness is balanced by the factor $M_\omega/\Gamma_\omega \approx 90$ from the $\omega$ propagator,
giving rise to an isospin-violating correction as large as 15\% on the amplitude level, corresponding to 
30\% in observables due to interference with the leading term.
Note also that the $\rho$--$\omega$ mixing amplitude has been pointed out to significantly enhance 
certain $CP$-violating asymmetries in hadronic $B$-meson decays~\cite{Gardner:1997yx}.

The effect of the  $\omega$ on the $\bar B_d^0 \to J/\psi \pi^+\pi^-$ decay can be related straightforwardly to
that on the pion vector form factor. To see this observe that the source
term for the $\pi\pi$ system is $\bar d d$ at tree level, see Fig.~\ref{fig:graph}, such that the isospin decomposition 
of the corresponding vector current reads
\be
\bar d\gamma^\mu d = -\frac{1}{2} \big( \bar u\gamma^\mu u - \bar d\gamma^\mu d\big)
+ \frac{1}{2} \big( \bar u\gamma^\mu u + \bar d\gamma^\mu d\big).
\ee
Comparison to Eq.~\eqref{eq:em-iso} shows that the relative strength of the isoscalar component differs from the 
electromagnetic current by a factor of $-3$, such that we will fix the $\rho$--$\omega$ mixing contribution 
in analogy to Eq.~\eqref{eq:omegamix}, but with the replacement 
$\kappa_{\rm em} \to \kappa = -3\kappa_{\rm em} \approx -5.4\times 10^{-3}$.
Notice that this is in contrast with the experimental analysis~\cite{Aaij:2014siy}, where the $\omega$
contribution is fitted with free coupling constants.

The (elastic) single-channel treatment, introduced in the beginning of this section, cannot be used in the $S$-wave case: there are strong inelastic effects in the region around 1~\GeV\ due to the opening of the $K\bar K$ channel, coinciding with the $f_0(980)$ resonance, which affects the phase of the scalar pion form factors (see e.g.\ the discussion in Ref.~\cite{Ananthanarayan:2004xy}). Thus the \Omnes problem has to be generalized, with the Watson theorem fulfilled in the elastic region and inelastic effects included above the $K\bar K$ threshold. This leads to the two-channel \MO equations that intertwine the pion and kaon form factors, 
defined as 
\begin{align} \label{eq:ffdef}
\left\langle \pi^+(p_1) \pi^-(p_2) \left|\bar qq\right|0\right\rangle &= \B^q\Gamma_\pi^q(s)  , \nonumber\\
\left\langle K^+(p_1) K^-(p_2) \left|\bar qq\right|0\right\rangle &= \B^q\Gamma_K^q(s) , 
\end{align}
where the quark flavors may be either $\bar qq=(\bar uu+\bar dd)/2$
for the light quarks, with the superscript $q=n$ denoting the
corresponding scalar form factor, or $\bar qq=\bar ss$ for strange
quarks (with superscript $q=s$).  Furthermore,
$\B^n=M_\pi^2/(m_u+m_d)$, $\B^s=(2M_K^2-M_\pi^2)/(2m_s)$.
Note that the form factors $\Gamma_{\pi,K}^q(s)$ are invariant under the QCD renormalization group, while the hadronic matrix elements are not due to the scale dependence inherent in the factors $\B^q$. This in turn allows for the cancellation of the scale dependence in the Wilson coefficients introduced in the effective Hamiltonian of Sec.~\ref{sec:amplconstr}.

Appealing to the tree-level diagram of Fig.~\ref{fig:graph}, 
we expect the non-strange scalar form factors to contribute dominantly in the $\bar B_d^0$ decay,
while the strange ones should feature mainly in the corresponding decay of the $\bar B_s^0$.
As discussed in detail below, these expectations are confirmed by the data analysis.

The \MO formalism is briefly reviewed in Appendix~\ref{app:MO}. 
It requires \textit{three} input functions: in addition to the $\pi\pi$ phase shift already necessary in the 
elastic case, modulus and phase of the $\pi\pi \to K\bar{K}$ $S$-wave amplitude also need to be known.
Our main solution is based on the Roy equation analysis by the Bern group~\cite{CCL2012,CCLprep} for the $\pi\pi$ phase shift,
the modulus of the $\pi\pi \to K\bar{K}$ $S$-wave as obtained from the solution of Roy--Steiner equations for 
$\pi K$ scattering performed in Orsay~\cite{BDM04}, and its phase from partial-wave analyses~\cite{Cohen, Etkin}.
Alternatively, we employ the $T$-matrix constructed by Dai and Pennington (DP) in Ref.~\cite{Dai:2014zta}:
here, a coupled-channel $K$-matrix parametrization is fitted to $\pi\pi$ data~\cite{pipidata1,pipidata2,pipidata3,pipidata4,pipidata5}, 
and the Madrid--Krak\'ow Roy-equation analysis~\cite{GarciaMartin:2011cn} is used as input; furthermore, 
the $K\bar K$ threshold region is improved by fitting also to Dalitz plot analyses of 
$D_s^+\to \pi^+\pi^-\pi^+$~\cite{Aubert:2008ao} and $D_s^+\to K^+K^-\pi^+$~\cite{delAmoSanchez:2010yp} by 
the BaBar Collaboration.

In addition, the channel coupling manifests itself through the fact that even in the simplest case, 
corresponding to the polynomial of Eq.~\eqref{eq:P.O} reducing to a constant, the scalar form factors
depend on \textit{two} such constants, corresponding to the form factor normalizations for both pion and kaon.
In contrast to the single-channel case, here the shape of the resulting form factors depends on the relative
size of these two normalization constants; on the other hand, once this relative strength is fixed, it relates the final states $\pi\pi$ and $K\bar K$ to each other unambiguously.  We will make use of this additional
predictiveness in Sec.~\ref{sec:KKprediction}.

In order to apply this formalism to the transversity partial waves we have to construct partial waves $f_\tau^{(\ell)}(s)$ that are free of kinematical singularities, i.e.\ represented by functions whose only non-analytic behavior is related to unitarity.
In Appendix~\ref{app:FFandPWE} the hadronic matrix element is introduced (using the basis of the momenta $p_\psi^\mu$, $P^\mu$, and $Q^\mu$, Eq.~\eqref{eq:PQ}) in terms of the form factors $\A_i$ and $\V$, Eq.~\eqref{eq:appMfi}, and related to the transversity basis, Eq.~\eqref{eq:AVF}.
Given that the form factors $\A_i$ and $\V$ are regular, Eq.~\eqref{eq:AVF} implies that there are additional factors of $X$, $\sigma_\pi$, and $\sqrt{s}$ introduced into the transversity form factors, which give rise to artificial branch cuts in the unphysical region. To avoid those, we write the partial waves as
\begin{align}
\F_0^{(S)} (s) &= X f_0^{(S)} (s),  &\F_0^{(P)} (s) &= \sigma_\pi f_0^{(P)} (s), \nonumber\\
\F_{\parallel}^{(P)}(s) &=  \sqrt{s} f_\parallel^{(P)} (s), 
&\F_{\perp}^{(P)}(s) &= \sqrt{s} X f_\perp^{(P)} (s),
\end{align}
where the $f_\tau^{(\ell)}$ are treated in the \Omnes formalism, i.e.\
\begin{equation}
f_0^{(S)}(s) = P_0^{(S,n)}(s) \Gamma_\pi^n (s) + P_0^{(S,s)}(s) \Gamma_\pi^s(s) , \qquad
f_\tau^{(P)}(s) = P_\tau^{(P)}(s)  \Omega_1^1 (s) .
\end{equation}
For the $S$-wave, we a priori allow for 
contributions of both non-strange ($n$) and strange ($s$) scalar form factors.
The coefficients of the polynomials $P_\tau^{(\ell)}(s)$ are to be determined from a fit to the efficiency-corrected and background-subtracted LHCb data, in particular to the angular moments $\langle Y _0 ^0 \rangle$ and $\langle Y _2 ^0 \rangle$.

Basically we assume the various polynomials to be well approximated by constants.
However, to study the impact of a linear correction at a later stage, we also consider linear polynomials $P_0^{(S,n)} = b^n_0 (1 + b'^n_0 s)$ and $P_\tau^{(P)} = a_\tau(1 + a'_\tau s)$ for the non-strange $S$-wave and the $P$-wave amplitudes, respectively.
The strange $S$-wave contribution is expected to be very small (in the LHCb analysis of $ \bar B_{d}^0 \to J/\psi \pi^+\pi^- $ the $f_0(980)$ meson is not seen), but tested in the fits. On the contrary, the $ \bar B_{s}^0 \to J/\psi \pi^+\pi^- $ distribution is dominated by the $f_0(980)$ resonance, described by a constant polynomial times \Omnes function, $P_0^{(S,s)} = c_0^s $, while there is no structure in the $f_0(500)$ region reported by LHCb. Thus in that case the non-strange $S$-wave amplitude is assumed to be negligible, to be confirmed in the fits.

Although the first $D$-wave resonance seen is the $f_2(1270)$, it may affect also the region below  $\sqrt{s} \approx 1~\GeV$ due to its finite width, $\Gamma_{f_2} = 185.1^{+2.9}_{-2.4}$~\MeV~\cite{Agashe:2014kda}. 
Therefore we also test its influence on the fit.
The $D$-waves could be treated in the same dispersive way as $S$- and $P$-waves, but this would increase the number of free parameters in our fits to the LHCb data. As the effect of $D$-wave corrections is rather small, we avoid introducing additional fit parameters and take over the amplitudes (with fixed couplings) used in the LHCb analysis, where the $f_2(1270)$ resonance is modeled by a Breit--Wigner shape. 

Since the data are given in arbitrary units, we collect all prefactors in normalizations that we subsume into the fit parameters (and into the transversity coefficients $\alpha_{\tau}^{f_2}$ that we extract from the LHCb fit results).
Writing  $ \langle Y _i ^0 \rangle$ in terms of \Omnes functions for $S$- and $P$-waves, supplemented by the $D$-wave resonance contribution, yields
{\allowdisplaybreaks
\begin{align}\label{eq:Yforfit}
\sqrt{4 \pi} \langle Y _0 ^0 \rangle &=  X \sigma_\pi \sqrt{s} \bigg\{ X^2 \big| b^n_0(1 + b'^n_0 s) \Gamma_\pi^n (s) + c_0^s \Gamma_\pi^s(s) \big|^2  \nonumber\\
&\quad  +  \sigma_\pi^2 \left| \Omega_1^1 (s) \right|^2 
\Big(  \big[ a_0 (1 + a'_0 s)\big]^2 + s \big[ a_\parallel ( 1 + a'_\parallel s)\big]^2 + s X^2 \big[ a_\perp ( 1 + a'_\perp s)\big]^2 \Big)
 \nonumber\\
&\quad  + \sum_{\tau =0, \bot, \| } \left| \alpha_\tau^{f_2} e^{i \phi_\tau^{f_2}} \A ^{(\tau)} _{f_2}(s) \right|^2 \bigg\}, \nonumber\\
\sqrt{4 \pi} \langle Y _2 ^0 \rangle &=  X \sigma_\pi \sqrt{s}\, \bigg\{
2 \Re \bigg( X \Big[ b^n_0 ( 1 + b'^n_0 s ) \Gamma_\pi^n (s) + c_0^s \Gamma_\pi^s(s)\Big]  \Big[\alpha_0^{f_2} e^{i \phi_0^{f_2}} \A ^{(0)}_{f_2}(s)\Big]^*\bigg) \nonumber\\
& \quad +
\frac{\sigma_\pi^2}{\sqrt{5}}  \left| \Omega_1^1 (s) \right|^2 \Big( 2 \big[a_0 (1 + a'_0 s)\big]^2  - s \big[a_\parallel (1+ a'_\parallel s)\big]^2 - s X^2 \big[a_\perp (1 + a'_\perp s)\big]^2 \Big)  \nonumber\\
 & \quad
  + \frac{\sqrt{5}}{7} \bigg(2 \Big| \alpha_0^{f_2} e^{i \phi_0^{f_2}} \A ^{(0)}_{f_2}(s) \Big|^2 + \sum_{\tau = \|, \bot }\Big| \alpha_{\tau}^{f_2} e^{i \phi_{\tau}^{f_2}}  \A_{f_2}^{(\tau)}(s) \Big|^2 \bigg)
\bigg\}.
\end{align}}%
For details concerning the definition of the Breit--Wigner amplitudes $\A_{f_2}^{(\tau)}(s)$, $\tau=0,\parallel,\perp$, see Ref.~\cite{Aaij:2014siy}.

\section{Fits to the LHCb data}\label{sec:fits}

\subsection[$\bar B_{d}^0 \to J/\psi \pi^+\pi^- $]{\boldmath{$\bar B_{d}^0 \to J/\psi \pi^+\pi^- $}}\label{sec:results}

We fit the angular moments $\langle Y _0 ^0 \rangle$ and $\langle Y _2 ^0 \rangle$, Eq.~\eqref{eq:Yforfit}, simultaneously. Taking up the discussion of Sec.~\ref{sec:omnes}, our basic fit, FIT~I, includes three fit parameters (to be compared to 14 free parameters in the Breit--Wigner parametrization used in the LHCb analysis, see below): the normalization factors for the $S$-wave ($b_0^n$)
and for two $P$-waves $f_{0}^{(P)}$ and $f_\parallel^{(P)}$ ($a_0, a_\parallel$). (We find that including the $\tau = \perp$ $P$-wave amplitude practically does not change the $\chi^2$, i.e.\ $a_\perp$ is a redundant parameter.) In the basic fit only $S$- and $P$-waves are considered. Beyond that, we study the relevance of certain corrections: in FIT~II we use again the same three parameters as in FIT~I, but in addition we include the $D$-wave contributions, fixed to their strengths as determined by LHCb. To further improve FIT~II, supplemental linear terms ($b'_0, a'_0, a'_\parallel$---cf.\ Eq.~\eqref{eq:Yforfit}) are allowed in FIT~III.
Performing FIT~III we find that two of the slope parameters, the linear non-strange $S$-wave term ($b'_0$) and the $\tau = \parallel$ $P$-wave slope ($a'_\parallel$), yield no significant improvement of the fits; their values are compatible with zero within uncertainties. 
We thus fix them to zero, and in FIT~III only the four parameters $b_0^n$, $a_0$, $a_\parallel$, and $a'_0$ are varied.
Furthermore, the effect of an inclusion of a strange $S$-wave component is tested. 
Its strength is found to be compatible with zero, justifying its omission. 

Note that the scalar pion form factors depend on the normalizations of both the pion and kaon form factors. While the normalizations in the case of the pion form factor are known quite precisely, there are considerable uncertainties for the kaon form factor normalizations, having an impact on the shapes of both pion form factors, see Appendix~\ref{app:MO}.
The non-strange kaon normalization $\Gamma^n_K(0)$ is limited to the range $(0.4 \dots 0.6)$. In our fits we fix the value to $\Gamma^n_K(0) = 0.5$, which is compatible with the current algebra result. The effect from a variation of $\Gamma^n_K(0)$ in the allowed interval shows up only in the second decimal place of the $\chi^2$/ndf.

\begin{figure}
\centering
\includegraphics[width=0.495\linewidth]{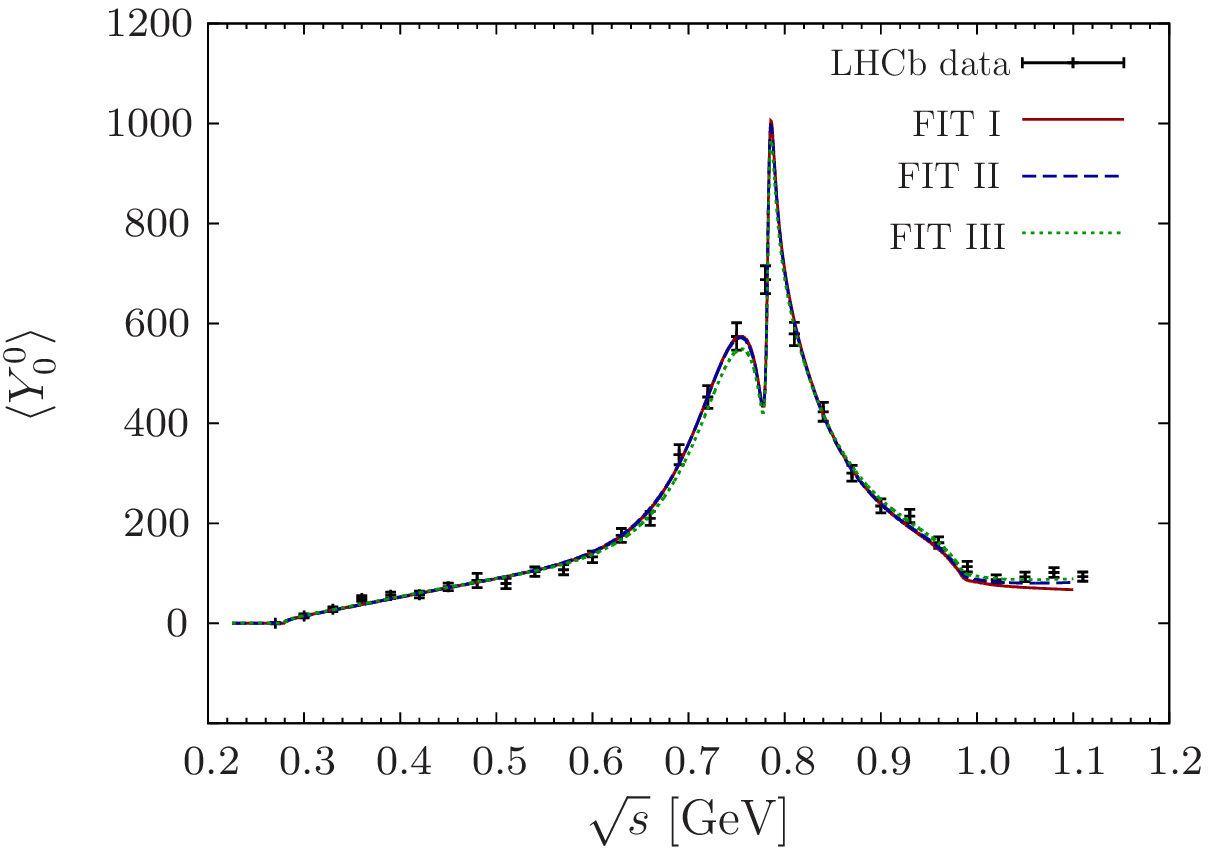}
\hfill
\includegraphics[width=0.495\linewidth]{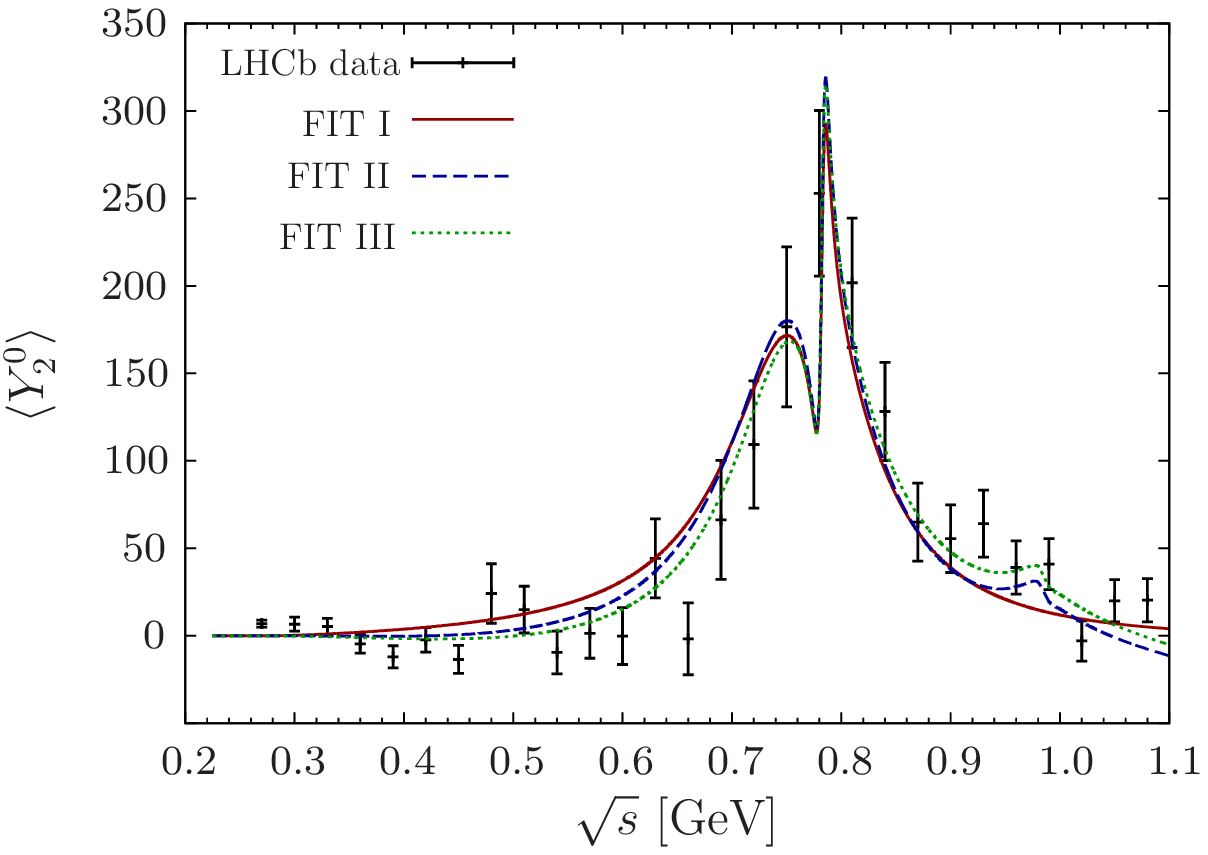}
\caption{$ \langle Y _0 ^0 \rangle$ (left) simultaneously fitted with $ \langle Y _2 ^0 \rangle$ (right), using 3 parameters without $D$-wave contribution (FIT~I, red, solid), and improving step by step by adding a Breit--Wigner-parametrized $D$-wave contribution (FIT~II, blue, dashed) and by allowing for 4 free parameters, also supplemented by the $D$-wave contribution (FIT~III, green, dotted).}
\label{fig:Y0i}
\end{figure}

The fitted coefficients and the resulting $\chi^2$/ndf, referring to Eq.~\eqref{eq:Yforfit}, are listed in Table~\ref{tab:fitYD}.
The large uncertainties can be traced back to the correlations between the fit parameters, especially present in FIT~III.
For a comparison to the LHCb fit, we insert their fit results (best model) into our definition of the $\chi^2$. In more specific terms this means that we do \emph{not} compare to the $\chi^2$ published in Ref.~\cite{Aaij:2014siy}, for which the full energy range up to $\sqrt{s}=2.1~\GeV$\ is fitted with 34 parameters and the data of all angular moments $\langle Y_i^0 \rangle $ for $i=0,\dots,5$ are included, but we calculate the $\chi^2$ in the region we use in our fits, i.e.\ including data up to $\sqrt{s}=1.02~\GeV$\ and the angular moments $\langle Y_0^0 \rangle $ and $\langle Y_2^0 \rangle $ only. We obtain $\chi^2_{\rm LHCb}/{\rm ndf} = 2.08$. In this limited energy range the Breit--Wigner description, including the $f_0(500)$, $\rho(770)$ and $\omega(782)$, requires 14 fit constants, while we have three (FIT~I, II) or four (FIT~III) free parameters and find $\chi^2/{\rm ndf} = 2.0$ (FIT~I), $\chi^2/{\rm ndf} = 1.5$ (FIT~II) and $\chi^2/{\rm ndf} = 1.3$ (FIT~III).
The calculated angular moments for the three fit models in comparison to the data are shown in Fig.~\ref{fig:Y0i}.

Probably the most striking feature of our solution is the pronounced effect of the $\omega$ that leads to the higher peak  in Fig.~\ref{fig:Y0i}.
As mentioned above, this isospin-violating contribution is fixed completely from an analysis of the pion vector form factor, however, its appearance
here is utterly different, since the coupling strength is multiplied by a factor of $-3$. This not only enhances the impact of the $\omega$ on the amplitude
level to about 50\%, but also implies that the change in phase of the signal is visible a lot more clearly: while in case of the vector form factor 
the $\omega$ amplitude leads to an enhancement on the $\rho$-peak and some depletion on the right wing, forming a moderate distortion of
the line shape, here we obtain a depletion on the $\rho$-peak accompanied by an enhancement on the right wing.
While the current data do not show the $\omega$ peak clearly, a small shape variation due to the $\rho$--$\omega$ interference is better seen in Ref.~\cite{Aaij:2014vda}, where a finer binning is used. 
The $\rho$--$\omega$ mixing strength obtained from a fit in that reference is consistent with the strength we obtain in a parameter-free manner.
Nonetheless, improved experimental data are called for, since an experimental
confirmation of the $\omega$ effect on $\bar B_d^0 \to J/\psi \pi^+\pi^-$ would allow one to establish that the $\bar B_d^0$ decay indeed provides a rather clean $\bar dd$ source.

\renewcommand{\arraystretch}{1.2}
\begin{table}
\centering
\begin{tabular}{@{}cccccc@{}}\toprule[1pt]
 & $\chi^2$/ndf & $|b_0^n|$ & $|a_0|$ & $|a_\parallel|$  & $a'_0$ \\
\midrule
FIT~I & 1.97 & $10.3_{-1.8}^{+1.5}$  & $46.5_{-6.8}^{+6.0}$ & $51.8_{-11.0}^{+9.0}$ & -- \\
FIT~II & 1.54 & $10.3_{-1.8}^{+1.5}$ & $47.6_{-6.6}^{+5.8}$ & $49.5_{-11.7}^{+9.4}$ & -- \\
FIT~III & 1.32 & $10.6_{-1.8}^{+1.5}$ & $37.7_{-21.3}^{+20.3}$ & $48.2_{-12.4}^{+9.8}$ & $0.4_{-0.7}^{+2.4}$ \\
\bottomrule
\end{tabular}
\caption{Resulting fit parameters and $\chi^2$/ndf for the various fit configurations FIT~I--III for the $ \bar B_{d}^0 \to J/\psi \pi^+\pi^- $ decay.}
\label{tab:fitYD}
\end{table}

\begin{figure}
\centering
\includegraphics[width=0.495\linewidth]{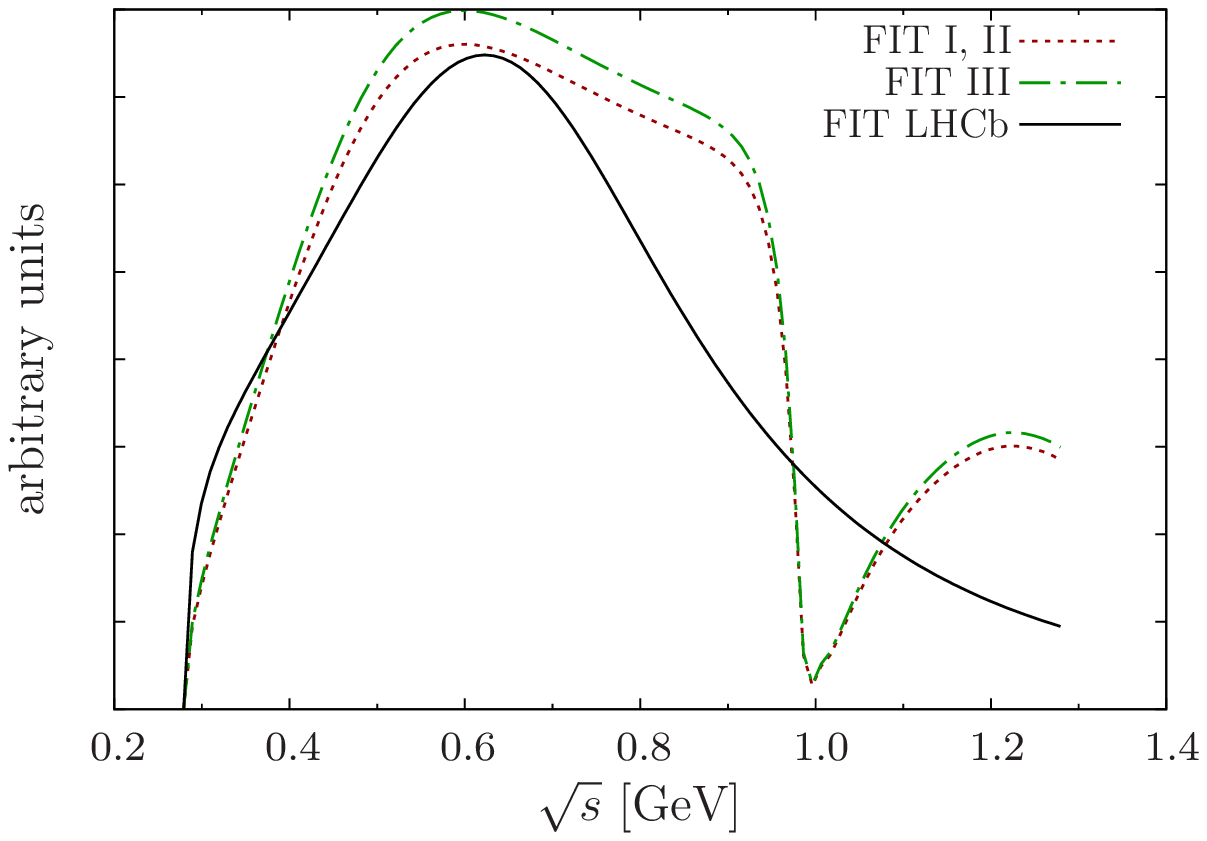}
\hfill
\includegraphics[width=0.495\linewidth]{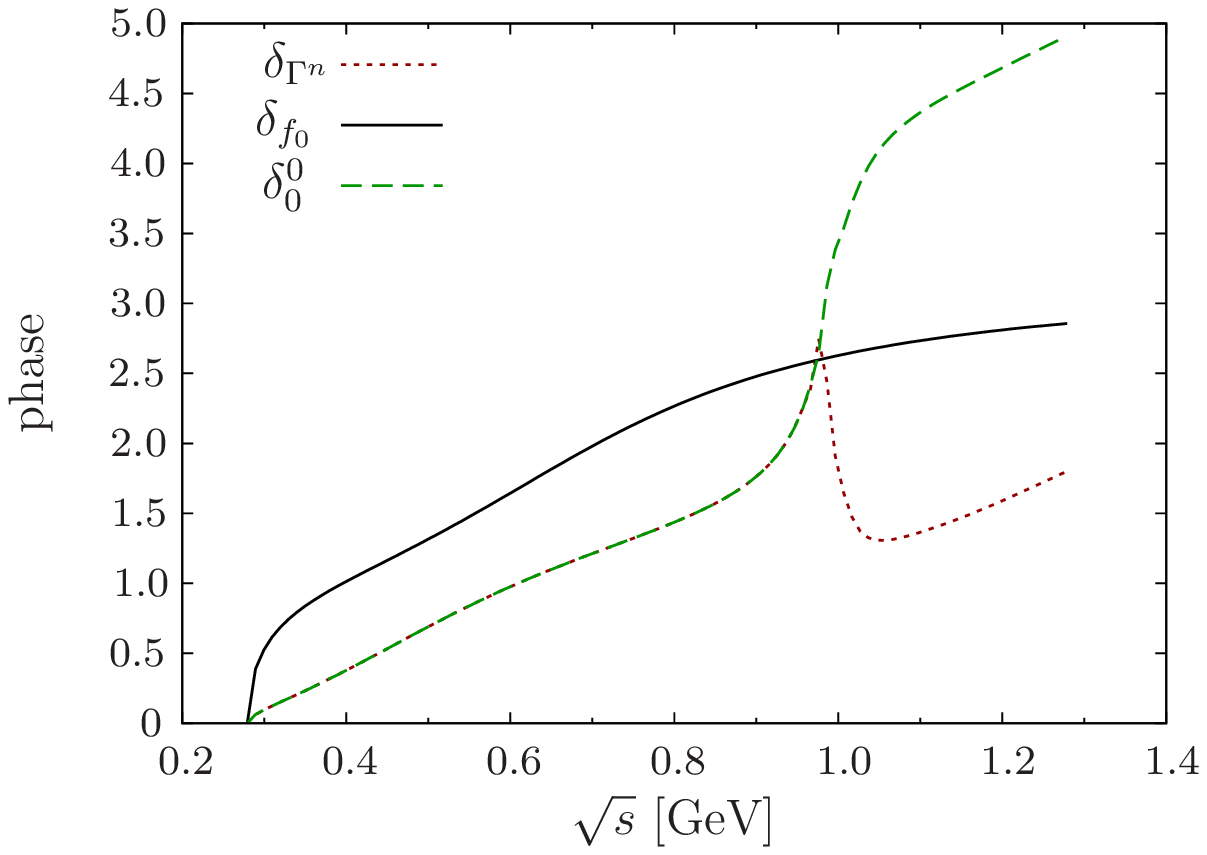}
\caption{Comparison of the $S$-wave amplitude strength and phase obtained in the LHCb and in our fits, respectively. In the left panel the $S$-wave part of the decay rate for the three fit configurations FIT~I--III is depicted together with the LHCb outcome. The right panel shows the phases of the non-strange scalar form factor $\delta_{\Gamma^n}$ (equal to the $\pi\pi$ $S$-wave phase shift $\delta_0^0$ below the $K \bar K$ threshold) compared to the $S$-wave phase $\delta_{f_0}$ extracted from the LHCb analysis.}
\label{fig:BdScompare}
\end{figure}

A key feature of the formalism employed here is its correct description of the $S$-wave. Figure~\ref{fig:BdScompare} shows the comparison of the $S$-wave amplitude strength of the LHCb Breit--Wigner parametrization with the ones obtained in FIT I--III, as well as the comparison of the corresponding phases.
In the elastic region, the phase of the non-strange scalar form factor $\delta_{\Gamma^n}= \text{arg}(\Gamma_\pi^n)$ coincides with the $\pi \pi$ phase shift $\delta_0^0$ that we use as input for the \Omnes matrix, in accordance with Watson's theorem. Right above the $K \bar K$ threshold, $\delta_{\Gamma^n}$ drops quickly, which causes the dip in the region of the $f_0(980)$, visible in the modulus of the amplitudes as well as the non-Breit--Wigner bump structure in the $f_0(500)$ region.
We find that the phase due to a Breit--Wigner parametrization largely differs from the dispersive solution, indicating that parametrizations of such kind
are not well suited for studies of $CP$ violation in heavy-meson decays.

Note that in the analysis of Ref.~\cite{Aaij:2014vda} the $f_0(500)$ is modeled not by a Breit--Wigner function, but by the theoretically better motivated parametrization of Ref.~\cite{Bugg:2006gc}. In this work, higher resonances are included by
multiplying $S$-matrix elements. While this procedure preserves unitarity, it produces terms at odds with any microscopic description of the coupled $\pi\pi$--$K\bar K$ system. As such also this approach introduces
uncontrolled theoretical uncertainties into the analysis. The only stringently model-independent way to include hadronic final-state interactions is via dispersion theory.

\subsection[$\bar B_{s}^0 \to J/\psi \pi^+\pi^- $]{\boldmath{$\bar B_{s}^0 \to J/\psi \pi^+\pi^-$}}\label{sec:results2}

The $ \bar B_{s}^0 \to J/\psi \pi^+\pi^- $ distribution in the region up to roughly 1~\GeV\ is clearly dominated by the $f_0(980)$. We therefore describe the data with the strange $S$-wave component only, using a constant subtraction polynomial ($c_0^s$). The only non-zero contribution to the fit thus comes from $\langle Y _0 ^0 \rangle$. Fitting the data up to $\sqrt{s} = 1.05~(1.02)~\GeV$\ yields $\chi^2/{\rm ndf} = 2.2~(1.8$) and $c_0^s = 16.8 \pm 0.4 ~(16.8 \pm 0.4$). 
In analogy to the $ \bar B_{d}^0$ decay we also perform the fit including the $D$-wave parametrization of the LHCb analysis. This yields an additional non-zero contribution to $\langle Y _2 ^0 \rangle$ due to the $S$--$D$-wave interference, which is fitted simultaneously with $\langle Y _0 ^0 \rangle$.  
Further, the influence of a linear subtraction polynomial for the strange $S$-wave is tested. However, none of these corrections exhibits a considerable improvement.

In the LHCb analysis the full energy range, $\sqrt{s} \le 2.1~\GeV$, is fitted with 22 (24) parameters for Solution I (II). Confining to the region we examine in our fit and considering the $f_0(980)$ resonance only, the number of fit parameters reduces to four (six), and we calculate $\chi^2_{\rm LHCb}/{\rm ndf} = 0.76 \, (0.82)$, when using our definition of the $\chi^2$.

\begin{figure}
\centering
\includegraphics[width=0.495\linewidth]{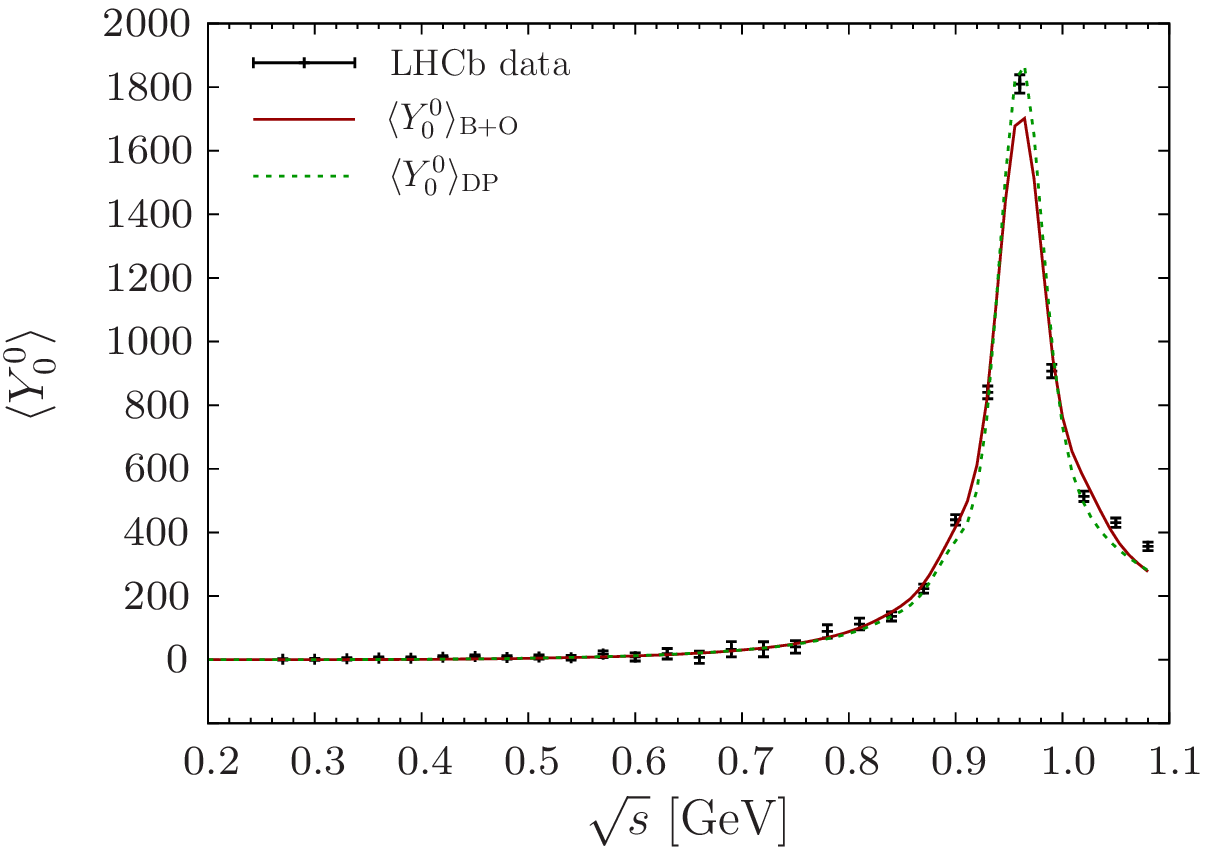}
\hfill
\includegraphics[width=0.495\linewidth]{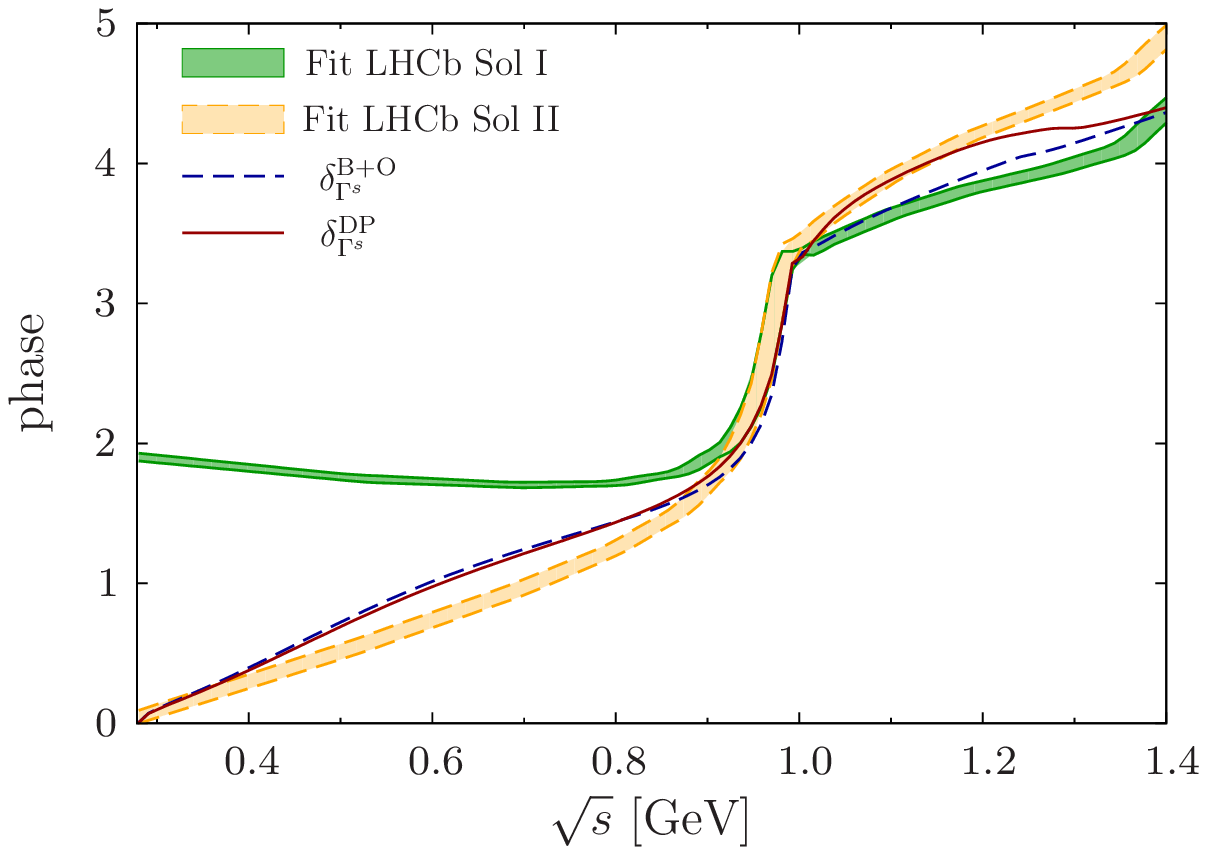}
\caption{Left panel: $ \langle Y _0 ^0 \rangle$ fitted using the strange $S$-wave with constant subtraction polynomial for two different phase inputs (red, solid: B+O input~\cite{CCL2012,CCLprep,BDM04}, green, dotted: DP input~\cite{Dai:2014zta}, based on the Madrid--Krak\'ow analysis~\cite{GarciaMartin:2011cn}).
Right panel: comparison of the phase of the strange scalar pion form factor for the B+O (blue, dashed) and DP (red, solid) input, respectively, with the $S$-wave phase extracted from the LHCb analysis (Solution~I and II, shown with error bands). 
}
\label{fig:Y0iBs}
\end{figure}

The strange scalar form factor, or the $f_0(980)$ peak in the dispersive formalism, depends crucially
on the $\pi\pi\to K \bar K$ $S$-wave transition amplitude, which is not as accurately known
as elastic $\pi\pi$ scattering (and even contains subtleties as non-negligible isospin breaking effects
due to the different thresholds of charged and neutral kaons, see e.g.\ Ref.~\cite{a0f0}).
As there are no error bands available for the \Omnes matrix (or the various input quantities),
to estimate the theoretical uncertainty we use and compare the fits resulting from the two different 
coupled-channel $T$-matrices described in Sec.~\ref{sec:omnes}.
A minimization of the $\chi^2$ using the modified \Omnes solution based on Ref.~\cite{Dai:2014zta} yields $\chi^2/{\rm ndf} = 3.4~(2.4$) and $c_0^s = 18.3 \pm 0.5~(18.2\pm 0.5$).\footnote{A similar procedure for the $\bar B_d^0$ decay has a rather small effect since the $S$-wave is not dominant in that case, and the difference of the $P$-wave phase of Refs.~\cite{GarciaMartin:2011cn,CCL2012,CCLprep} is quite small (the $S$- or $P$-wave phase modification yields, in the most perceptible cases, a 4\% correction of the $\chi^2$).}
The resulting $\langle Y _0 ^0 \rangle$ curves for both fits, using the phase input from the Bern~\cite{CCL2012,CCLprep}
and Orsay~\cite{BDM04} groups (B+O),
as well the one of Ref.~\cite{Dai:2014zta} (DP), are presented in Fig.~\ref{fig:Y0iBs}.
Furthermore we show the phase shifts and the phases of the strange form factor for both phase inputs and  compare to the LHCb phase due to Solution~II 
(with $f_0(980)$ and a non-resonant $S$-wave contribution) as well as Solution~I ($f_0(980)$ parametrization only). While the latter phase has a negative slope for $s \lesssim 1~\GeV$, which does not agree with the known phase shift, the phase extracted in Solution~II is remarkably close to both the Bern and Madrid phase motions.

\subsection[$\bar B_{s}^0 \to J/\psi K^+ K^-$ $S$-wave prediction]{\boldmath{$\bar B_{s}^0 \to J/\psi K^+ K^-$ $S$}-wave prediction}\label{sec:KKprediction}

Having obtained the $\bar B_{s}^0 \to J/\psi \pi^+ \pi^-$ fit parameters, we can make a prediction for the $\bar B_{s}^0 \to J/\psi K^+ K^-$ $S$-wave amplitudes, using the relation between the $\pi\pi$ and the $K \bar K$ final states provided by the coupled-channel formalism, cf.\ Appendix~\ref{app:MO}.\footnote{In the case of the $\bar B_{d}^0 \to J/\psi K^+ K^-$ decay~\cite{Aaij:2013mtm}, the prediction of the $S$-wave does not work in such a direct way due to the $I=1$ $S$-wave contribution (with a prominent $a_0(980)$ resonance) in addition to $f_0$ resonances in the $I=0$ $S$-wave.} 

In particular an understanding of the $S$-wave background to the prominent $\phi(1020)$ is of interest. 
In the LHCb analysis~\cite{Aaij:2013orb}, the $f_0(980)$ as well as a non-resonant $S$-wave content is reported within a mass window of $\pm 12$~\MeV\ around the $\phi(1020)$, which contribute an $S$-wave fraction of $(1.1 \pm 0.1^{+0.2}_{-0.1})\%$---consistent with former measurements from LHCb, CDF, and ATLAS~\cite{lhcbconf2012, Aaltonen:2012ie, Aad:2012kba}, as well as theoretical  estimates~\cite{Stone:2008ak}. 
We calculate the $S$-wave fraction in the same mass interval $\pm 12$~\MeV\ around the $\phi(1020)$ mass adopting the LHCb Breit--Wigner parametrization for the $\phi(1020)$, but using the predicted $S$-wave for the $J/\psi K^+ K^-$ final state. 
Naively, this $S$-wave can be obtained by replacing the pion scalar form factor and all pion masses and momenta by the respective kaon quantities and taking the resulting fit parameters from the pion case. However, the fit result depends on the normalization of the $\bar B_{s}^0 \to J/\psi \pi^+\pi^-$ distribution. Hence, taking over the pion fit results for such a prediction requires a proper normalization of both decay channels relative to each other. To achieve this, we use the absolute branching fractions~\cite{Agashe:2014kda}
\bea
\B\left(\bar B_{s}^0 \to J/\psi K^+  K^-\right) &=& (7.9 \pm 0.7) \times 10^{-4},\nonumber\\
\B\left(\bar B_{s}^0 \to J/\psi \pi^+\pi^-\right)&=&(2.12\pm 0.19) \times 10^{-4}, \nonumber
\eea
and define normalization constants
\be
\N_{\{\pi,K\}} = \frac{\B(\bar B_{s}^0 \to J/\psi \{\pi^+\pi^-, K^+ K^-\})}{N(\bar B_{s}^0 \to J/\psi \{\pi^+\pi^-, K^+ K^-\})}\,,
\ee
where
\be
N(\bar B_{s}^0 \to J/\psi \{\pi^+\pi^-, K^+ K^- \} ) = \sqrt{4\pi} \int \left\langle Y_0^0\left(\bar B_{s}^0 \to J/\psi \{\pi^+\pi^-, K^+ K^- \} \right)\right\rangle \diff\sqrt{s}
\ee
is the total number of events.\footnote{For the $\bar B_{s}^0 \to J/\psi K^+ K^-$ decay~\cite{Aaij:2013orb} no data for the efficiency-corrected angular moments are available. We therefore extract the strength of the $\phi(1020)$ Breit--Wigner amplitude from the published expected signal yield $N_{\rm exp}$ and use $N(\bar B_{s}^0 \to J/\psi K^+ K^-) = \B(\bar B_{s}^0 \to J/\psi K^+ K^-) / N_{\rm exp}$.}

The $S$-wave contribution to the $\phi (1020)$ peak region is given by
\be 
\R_{S/\phi} \equiv \frac{\N_\pi \int_{m_\phi - 12\,\MeV}^{m_\phi + 12\,\MeV} X^3 \sigma_K \sqrt{s} \left| c_0^s \Gamma_K^s(\sqrt{s}) \right|^2 \, \diff\sqrt{s}}{
\ \int_{m_\phi - 12\,\MeV}^{m_\phi + 12\,\MeV} \sqrt{4\pi}\big\langle \tilde Y_0^0\left(\bar B_{s}^0 \to J/\psi  K^+ K^- \right)\big\rangle  \, \diff\sqrt{s}},
\ee
where we can approximate the (normalized) angular moment in the region of interest by the $S$-wave and the $\phi(1020)$ contribution,
\bea
&&\left.\sqrt{4\pi}\left\langle \tilde Y_0^0\left(\bar B_{s}^0 \to J/\psi  K^+ K^- \right)\right\rangle\right|_{|\sqrt{s}-m_\phi| \lesssim 12~\MeV}\nonumber\\
&& \qquad\qquad \approx
X \sigma_K \sqrt{s} \bigg(X^2
\N_\pi \left| c_0^s \Gamma_K^s(\sqrt{s}) \right|^2 + \N_K
\sum_\tau \left|  
\alpha_{\tau}^{\phi} \A_{\phi}^{(\tau)}(s)
\right|^2
\bigg).
\eea
Using the B+O input, we obtain $\R_{S/\phi} = 1.1\%$, in agreement with the LHCb result. 
However, there is a notable uncertainty due to 
the estimated ambiguity in the phase input in the region of the $f_0(980)$ resonance discussed in Sec.~\ref{sec:results2}. Using the DP phase instead of the B+O phase input yields a fraction of $1.95\%.$

\section{Summary and outlook}\label{sec:summary}

In this article, we have described the strong-interaction part of the $\bar B_d^0 \to J/\psi \pi^+\pi^- $ and $\bar B_s^0 \to J/\psi \pi^+\pi^-$ decays by means of dispersively constructed scalar and vector pion form factors.
This formalism respects all constraints from analyticity and unitarity.
The non-strange and strange scalar form factors are calculated from a two-channel \MO formalism that 
requires the pion--pion elastic $S$-wave phase shift as well as modulus and phase of the corresponding
$\pi\pi\to K\bar K$ amplitude as input.  
For the vector form factor, an elastic \Omnes representation based solely on the pion--pion $P$-wave phase shift
is sufficient, supplemented by an enhanced isospin-breaking contribution of $\rho$--$\omega$ mixing, which can be 
fixed from data on $e^+e^-\to\pi^+\pi^-$.  

For energies $\sqrt{s} \le 1.02~\GeV$, a minimal description of all $S$- and $P$-waves
(constructed in a form free of kinematical singularities) as the corresponding form factors, 
multiplied by real constants, has been shown to be sufficient.  Allowing for subtraction polynomials
with linear $s$-dependence leads to a slightly improved fit quality 
solely in the case of one $P$-wave component, with a slope still compatible 
with zero within uncertainties.  
In particular considering the $S$-wave slope as a free fit parameter (as opposed to fixing it to zero) 
only yields a minimal improvement of the $\chi^2$.
In accordance with expectations from the underlying tree-level decay mechanism, below the 
onset of $D$-wave contributions that become important with the $f_2(1270)$,
only the non-strange scalar and the vector form factors feature in the $\bar B_d^0$ decay, 
while the strange scalar form factor determines the $\bar B_s^0$ $S$-wave.

The overall fit quality in the energy range considered is at least as good as in the 
phenomenological fits by the LHCb collaboration~\cite{Aaij:2014siy,Aaij:2014emv}, where Breit--Wigner resonances and
non-resonant background terms were used. However, since the dispersive
analysis allows one to use input from other sources, our analysis calls for a much smaller number
of parameters to be determined from the data. In addition,
a comparison of the $\bar B_d^0$ $S$-wave obtained from the dispersive analysis with the one deduced from the LHCb analysis 
shows drastic differences in both modulus and phase:
it is well-known that the $f_0(500)$ does not have a Breit--Wigner shape, and therefore such parametrizations should be 
avoided---especially when it comes to studies of $CP$ violation that need a reliable treatment of the phases induced by the hadronic
final-state interactions~\cite{Gardner:2001gc}.
The LHCb analysis of the $\bar B_s^0$ $S$-wave uses a Flatté parametrization of the $f_0(980)$, solely (corresponding to their Solution~I) or combined with a non-resonant background (Solution~II).  
Only Solution~II yields a phase that is close to the phase of the strange scalar form factor,
and approximately compatible with Watson's final-state interaction theorem in the elastic region.

Finally we have made a prediction for the $\bar B_s^0 \to J/\psi K^+K^- $  $S$-wave, 
which is related to the corresponding $\pi^+\pi^-$ final state through channel coupling.
Only the results of the fit to the $\pi^+\pi^-$ final state are required to predict 
an $S$-wave fraction below the $\phi(1020)$ resonance of about 1.1\%, in agreement 
with the findings by the LHCb collaboration.
We have not attempted a corresponding prediction for the $\bar B_d^0 \to J/\psi K^+K^- $ $S$-wave,
since this has an isovector component (corresponding e.g.\ to the $a_0(980)$ resonance).
This would have to be described by a coupled-channel treatment of the $\pi\eta$ and $K\bar K$ 
$S$-waves~\cite{Albaladejo:2015aca}.

To extend our description of the form factors to higher energies, eventually covering most of the
energy range accessible in $\bar B_{d/s}^0 \to J/\psi \pi^+\pi^-$,
inelastic channels with corresponding higher resonances have to be taken into account. 
Here, a formalism recently developed for the vector form factor~\cite{Hanhart:2012wi} that correctly
implements the analytic structure and unitarity, reduces to the \Omnes representation
in the elastic regime, but maps smoothly onto an isobar-model picture at 
higher energies should be extended to the scalar sector.  
Even an \textit{extraction} of the scalar form factors from these high-precision LHCb data sets 
seems feasible, and should be pursued in the future.

\acknowledgments
We would like to thank the LHCb collaboration for the invitation to the Amplitude Analysis Workshop
where this work was initiated, and in particular Tim Gershon, Jonas Rademacker, Sheldon Stone,
and Liming Zhang for useful discussions.
We are furthermore grateful to Mike Pennington for providing us with the coupled-channel $T$-matrix parametrization 
of Ref.~\cite{Dai:2014zta}.
Financial support by DFG and NSFC through funds provided to the Sino--German CRC~110 
``Symmetries and the Emergence of Structure in QCD''
is gratefully acknowledged.

\appendix

\section{Form factors and partial-wave expansion}\label{app:FFandPWE}

In the standard basis of momenta $p_\psi$, $P^\mu$, and $Q^\mu$, Eq.~\eqref{eq:PQ}, the matrix element describing the hadronic part of the $\bar B_{d/s}^0$ decay is given by four dimensionless form factors, three axial ($\A_i$) and one vector ($\V$),
\begin{align}\label{eq:appMfi}
\M^{\pi\pi}_\mu &= \langle \pi^+(p_1) \pi^-(p_2) | J_\mu^{d/s} | \Bd (p_B) \rangle
= P_{\mu} \A_1 + Q_{\mu} \A_2 + (p_{\psi})_\mu \A_3 + i \epsilon_{\mu\nu\rho\sigma} p_\psi^\nu P^\rho Q^\sigma \V,\nonumber\\
J_\mu^{q} &= \bar{q} \gamma_\mu (1-\gamma_5) b.
\end{align}

In Sec.~\ref{sec:matrixelement} we use a different (orthogonal) basis of momentum vectors, $p_\psi^\mu$,
$\bar p_{(\perp)}^\mu$, and $Q_{(\parallel)}^\mu$, see Eq.~\eqref{eq:newbasis}, 
corresponding to the orthonormal basis of polarization vectors of the $J/\psi$ meson~\cite{Faller:2013dwa},
\be
\epsilon^\mu (t) = \frac{p_\psi^\mu}{M_\psi}, \quad 
\epsilon^\mu (0) = -\frac{M_\psi}{X} P^\mu_{(0)}, \quad 
\epsilon^\mu (\pm) = -\frac{1}{\sqrt{2s}\, \sigma_\pi \sin \theta_\pi} \left(Q_{(\parallel)}^\mu \mp i \bar p_{(\perp)}^\mu \right) e^{\mp i \phi}.
\ee
This allows us to describe the matrix element $\M_{\pi\pi}^\mu$ in terms of the transversity form factors, Eq.~\eqref{eq:Mfi},
or similarly (with regard to an easily performable partial-wave expansion) in terms of helicity form factors, defined via the contraction of $\M_{\pi\pi}^\mu$ with the polarization vector,
\be
\mathH_\lambda = \langle \pi^+ \pi^- |  J_\mu^{d/s} | \Bd  \rangle \, \epsilon_\mu^\dagger (\lambda).
\ee
The relations between the transversity and helicity form factors can be read off to be
\be\label{eq:appHF}
\mathH_t = \F_t, \quad \mathH_0 = \F_0, \quad \mathH_\pm = (\F_\parallel \pm \F_\perp) \frac{\sigma_\pi}{\sqrt{2}} \sin \theta_\pi e^{\pm i \phi},
\ee
as well as those to the set \{$\A_i, \, \V$\},
\bea\label{eq:AVF}
\F_\perp &=& -\sqrt{s} X \V , \quad  \F_\parallel = \sqrt{s} \A_2, \quad \F_0 = \frac{X}{2M_\psi} \left(2 \A_1 + \frac{\sigma_\pi \cos \theta_\pi (P \cdot p_\psi)}{X} \A_2 \right), \nonumber\\
\F_t &=&  \frac{P \cdot Q}{M_\psi} \A_1 + \frac{\sigma_\pi X \cos \theta_\pi}{2 X M_\psi} \left( (P \cdot p_\psi)(P \cdot Q) - s M_\psi^2 \right) \A_2 + M_\psi \A_3.
\eea
The \textit{unphysical} time component $\F_t$ does not contribute. We expand the remaining three form factors $\mathH_{0,\pm}$ in partial waves.
The latter relation is of particular interest when defining partial waves that are free of kinematical singularities and zeros, see Sec.~\ref{sec:omnes}.

The partial-wave expansion of the helicity amplitudes reads
\be
\mathH_\lambda (s) = \sum_\ell \sqrt{2\ell +1} \mathH_\lambda^{(\ell)} (s) d_{\lambda 0}^\ell (\theta_\pi) e^{\lambda i \phi},
\ee
where the $d_{\lambda \lambda'}^\ell$ are the small Wigner-$d$ functions. 
Using
\be
d_{00}^\ell (\theta_\pi) = P_\ell (\cos \theta_\pi), \quad d_{10}^\ell (\theta_\pi) = - d_{-1 0}^\ell (\theta_\pi) = -\frac{\sin \theta_\pi}{\sqrt{\ell (\ell+1)}} P_\ell'(\cos\theta_\pi), 
\ee
we see that the zero-component $\mathH_0(s)$ is expanded in terms of Legendre polynomials $P_\ell (\cos \theta_\pi) $ and thus contains all $S$-, $P$-, and $D$-wave contributions, while the $\mathH_\pm(s)$ partial-wave expansions, proceeding in derivatives of the Legendre polynomials $P_\ell'(\cos\theta_\pi)$, start with the $P$-wave amplitudes, i.e.\
\bea
\mathH_0 (s) &=& \mathH_0^{(S)} (s) + \sqrt{3} \cos\theta_\pi \mathH_0^{(P)} (s) + \frac{\sqrt{5}}{2} \left(3 \cos^2\theta_\pi - 1 \right) \mathH_0^{(D)} (s) + \dots \,,\nonumber\\
\mathH_\pm (s) &=& \mp \sqrt{\frac{3}{2}} \sin \theta_\pi \left( \mathH_\pm^{(P)} (s) + \sqrt{5} \cos\theta_\pi \mathH_\pm^{(D)} (s) \right)  e^{\pm i \phi} + \dots ,
\eea
where the ellipses denote $F$-waves and larger.
Equivalently, due to Eq.~\eqref{eq:appHF} and using $\mathH_\pm^{(\ell)} (s) = \mp  \frac{\sigma_\pi}{\sqrt{2}} \big( \F_\parallel^{(\ell)}(s) \pm  \F_\perp^{(\ell)}(s)\big)$, we arrive at the partial-wave expansion of the transversity form factors given in 
Eq.~\eqref{eq:F-PWE} in the main text.

In order to calculate the differential decay rate we sum over the squared helicity amplitudes, 
\be
\left| \overline \M \right|^2 = \frac{G_F^2}{2} |V_{cb}|^2 |V_{cq}|^2 f_\psi^2 M_\psi^2 \left( |\mathH_0 |^2 + |\mathH_+ |^2 +|\mathH_- |^2 \right) \quad (q=\{d, s\})
\ee
and integrate over the invariant three-particle phase space, which is given by
\be
\diff \Phi^{(3)} = \frac{X \sigma_\pi}{4 (4\pi)^2 m_B^2}\, \diff s\, \diff\cos\theta_\pi \,\diff\phi.
\ee
Neglecting waves larger than $D$-waves and integrating over $\phi$ we arrive at Eq.~\eqref{eq:d2Gamma}.

\section{Coupled-channel \Omnes formalism}\label{app:MO}

We briefly discuss the coupled-channel derivation of the scalar pion and kaon form factors ($I=0, \ell=0$). 
The two-channel unitarity relation reads
\begin{equation} \label{eq:ImF=TSF}
\mbox{disc}\,\vec{\Gamma} (s) = 2 i T^{0*}_0(s) \Sigma (s) \vec{\Gamma} (s), 
\end{equation}
where the two-dimensional vector $\vec{\Gamma}(s)$ contains the pion and kaon scalar isoscalar form factors and $T^0_0(s)$ and $\Sigma(s)$ are two-dimensional matrices,
\begin{equation} \label{eq:T_2channel}
T^0_0(s) = \left( \begin{array}{cc}
\dfrac{\eta^0_0 (s) e^{2 i \delta^0_0 (s)} - 1}{2i {\sigma}_{\pi} (s)} & |g^0_0(s)| e^{i \psi^0_0 (s)}  \\[6pt]
|g^0_0(s)| e^{i \psi^0_0 (s)} & \dfrac{\eta^0_0 (s) e^{2 i (\psi^0_0 (s) - \delta^0_0 (s))} - 1}{2i {\sigma}_K (s)}  \end{array} \right),
\end{equation}
and $\Sigma(s)=\mbox{diag} \left(\sigma_{\pi} (s) \Theta(s-4M_{\pi}^2),\sigma_K (s) \Theta(s-4 M_K^2)\right)$, with $\sigma_i (s) = \left(1-4M_i^2/s\right)^{1/2}$ and $\Theta(.)$ denoting the Heaviside function.
There are \textit{three} input functions entering the $T$-matrix, the $\pi\pi$ $S$-wave isoscalar phase shift $\delta^0_0(s)$ and the $\pi\pi \to K\bar{K}$ $S$-wave amplitude $g^0_0(s)=|g^0_0(s)| {\rm exp}(i\psi^0_0(s))$ with modulus and phase. 
The modulus $|g^0_0(s)|$ is related to the inelasticity parameter $\eta^0_0(s)$ by
\begin{equation}
\eta^0_0 (s) = \sqrt{1-4\sigma_{\pi}(s) \sigma_K(s)|g^0_0(s)|^2\Theta(s-4M_K^2)}.
\end{equation}
Writing the two-dimensional dispersion integral over the discontinuity~\eqref{eq:ImF=TSF} leads to a system of coupled \MO equations,
\begin{equation} \label{eq:vecFdisp}
\vec{\Gamma} (s) = \frac{1}{\pi} \int_{4M_{\pi}^2}^{\infty} \frac{T^{0*}_0(s') \Sigma (s') \vec{\Gamma} (s')}{s'-s-i\epsilon}\diff s'.
\end{equation} 
A solution can be constructed introducing a two-dimensional \Omnes matrix, which is connected to the form factors by means of a multiplication with a vector containing the normalizations $\Gamma_{\pi} (0)$ and $\Gamma_K(0)$~\cite{DGL90},
\begin{equation}\label{eq:2channelMOs}
\left(
\begin{array}{c}
\Gamma_{\pi} (s) \\ \smash{\frac{2}{\sqrt{3}}} \Gamma_K(s) 
\end{array}
\right)
=
\left(
\begin{array}{cc}
\Omega_{11} (s) & \Omega_{12} (s)\\ \Omega_{21}(s) & \Omega_{22} (s)
\end{array}
\right) 
\left(
\begin{array}{c}
\Gamma_{\pi} (0) \\ \smash{\frac{2}{\sqrt{3}}}\Gamma_K(0) 
\end{array}
\right),
\end{equation}
where $\Gamma_{\pi,K}(s)$ represents both strange and non-strange form factors, $\Gamma^s_{\pi,K}(s)$ and $\Gamma^n_{\pi,K}(s)$,  which differ merely in their respective normalizations.
Thus the problem reduces to finding a matrix $\Omega(s)$ that fulfills
\be
\Im\,\Omega (s) = T^{0*}_0 (s) \Sigma (s) \Omega (s),\quad
\Omega (s) = \frac{1}{\pi} \int_{4M_{\pi}^2}^{\infty} \frac{T^{0*}_0 (s') \Sigma (s') \Omega (s')}{s'-s-i\epsilon}\diff s',\quad
\Omega (0) = \mathbbm{1},
\ee
which has to be solved numerically~\cite{DGL90, Moussallam2000, Sebastien, Hoferichter:2012wf}.
To ensure an adequate asymptotic behavior, we exploit the correlation between the high-energy behavior of the \Omnes solution and the sum of the eigen phase shifts $\sum \delta_\ell^I(s)$~\cite{Moussallam2000},
\be
\sum \delta_\ell^I(s) \overset{s\to \infty}{\longrightarrow} \,  m \pi = \left\{
\begin{array}{cc}
\pi \quad \;\,{\rm for} \quad I=1,~\ell=1\\
2 \pi \quad {\rm for} \quad I=0,~\ell=0
\end{array} \right. \quad \quad \Longleftrightarrow \quad \Omega_\ell^I(s)\overset{s\to \infty}{\longrightarrow}\, \frac{1}{s} \, ,
\ee
where $m$ is the number of channels that are treated in the formalism.

According to the Feynman--Hellmann theorem, the form factors for zero momentum are related to the corresponding Goldstone boson masses, which at next-to-leading order in the chiral expansion in terms of quark masses depend on certain low-energy constants. These are determined in lattice simulations with $N_f=2+1$ dynamical flavors at a running scale $\mu = \unit{770}{\MeV}$~\cite{Aoki:2013ldr}, limiting the form factor normalizations to the 
ranges\footnote{Similar ranges, with slightly increased values in the case of the kaon form factor normalizations, are found in simulations with $N_f=2+1+1$ dynamical flavors~\cite{Dowdall:2013rya}.}
\begin{align}\label{eq:FFNorm}
\Gamma_{\pi}^n(0) &= 0.984 \pm 0.006, & \Gamma_{\pi}^s(0) &= (-0.001 \dots 0.006)  \approx 0, \nonumber \\
\Gamma_K^n(0) &=(0.4 \dots 0.6), &  \Gamma_K^s(0) &= (0.95 \dots 1.15) .
\end{align}

\begin{figure}
\centering
\includegraphics[width=0.48\linewidth]{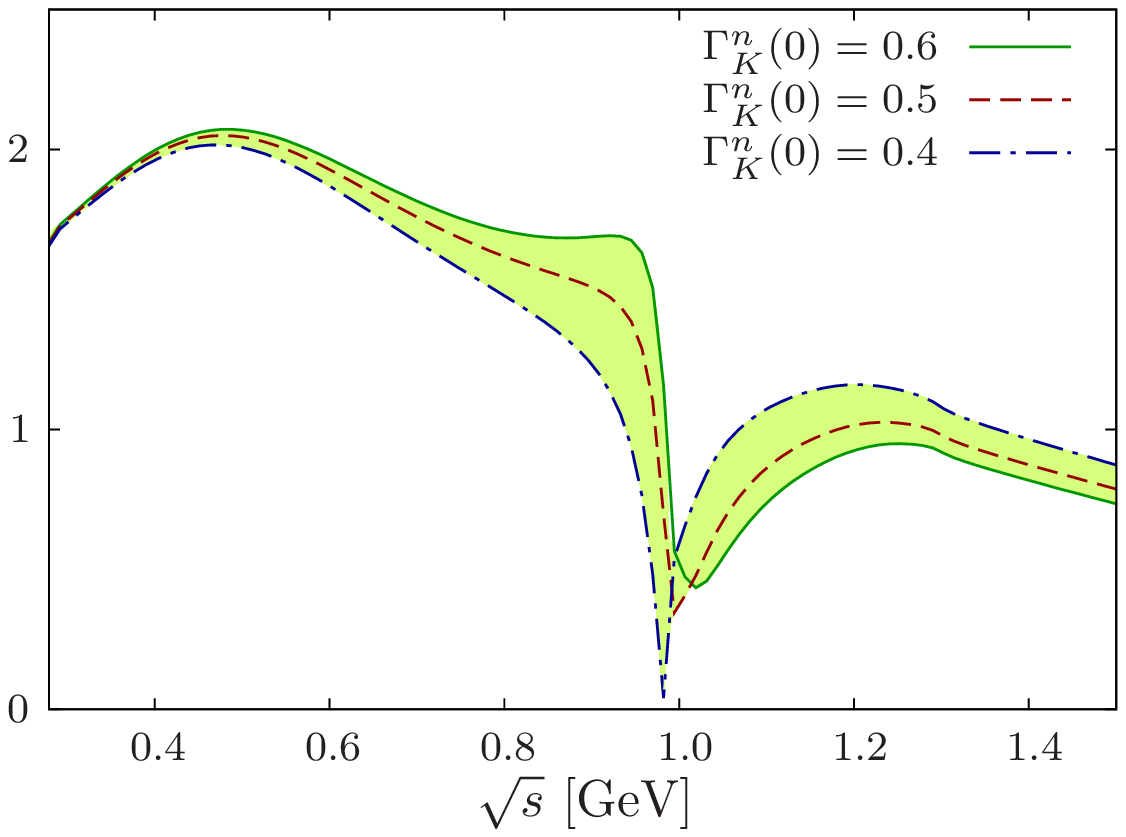}
\hfill
\includegraphics[width=0.48\linewidth]{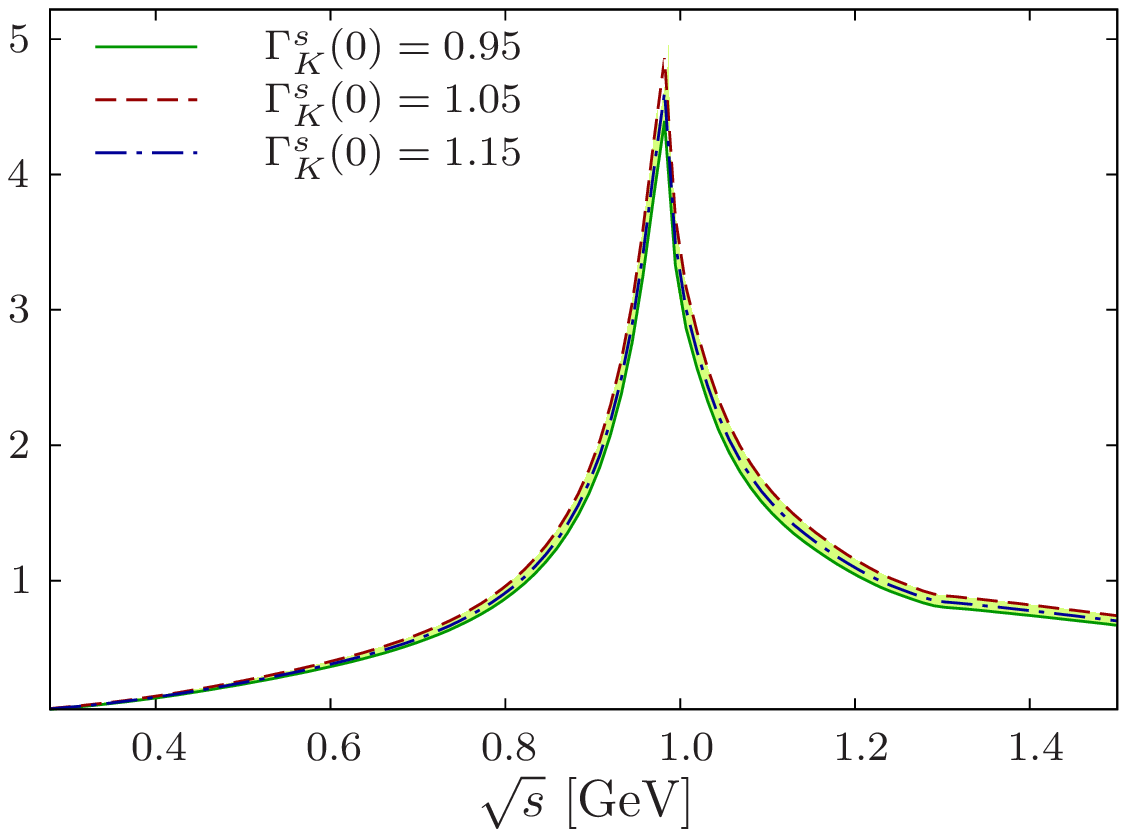}
\caption{Modulus of the scalar pion non-strange (left panel) and strange (right panel) form factors, depicted for three different normalizations inside the allowed range, illustrated by the uncertainty band.}
\label{fig:ScalarFF}
\end{figure}
Figure~\ref{fig:ScalarFF} shows the results obtained for the modulus of the pion form factor (see also Ref.~\cite{Daub:2012mu}). The sensitivity due to the uncertainty in the kaon form factor normalization is illustrated by the uncertainty bands.
The strange form factor exhibits a peak around $1~\GeV$, which is produced by the $f_0(980)$ resonance. On the contrary in the pion non-strange form factor the $\sigma$ meson appears as a broad bump (notice the non-Breit--Wigner shape) around 500~\MeV.

\bibliography{BJpsi}

\end{document}